\documentclass[journal=jacst,manuscript=article,layout=onecolumn]{achemso}

\usepackage{chemformula} % Formula subscripts using \ch{}
\usepackage[colorlinks=true, allcolors=black]{hyperref}

\usepackage[version=3]{mhchem} % Formula subscripts using \ce{}
\emergencystretch=\maxdimen
\usepackage{dsfont}
\usepackage{amssymb}
\usepackage{algorithm2e}
\usepackage{xcolor}
\RestyleAlgo{ruled}
\SectionNumbersOn
\usepackage{chngcntr} % for counting SI figures/tables

\mciteErrorOnUnknownfalse  

\newcommand{\cfg}{x}
\newcommand{\img}{y}
\newcommand{\ImgSet}{Y}

\newcommand{\Npix}{{N_{\text{px}}}}
\newcommand{\Natom}{{N_{\mathrm{atom}}}}
\newcommand{\Npoints}{T}
\newcommand{\Ncenclusters}{M}
\newcommand{\ccenter}{\chi}  % Center of clusters
\newcommand{\weight}{\alpha}
\newcommand{\truedensity}{\rho}  % True density (i.e. Boltzmann).
\newcommand{\rwt}{\eta}  % Reweighting function for MD
\newcommand{\pguess}{{p_0}}  % Partition fxn of the Hamiltonian that we are correcting.

\newcommand{\1}{\mathds{1}}  % Indicator Function

\author{Wai~Shing~Tang}
\affiliation[Flatiron Institute -CCM]
{Center for Computational Mathematics, Flatiron Institute, New York, USA}
\alsoaffiliation[Flatiron Institute -CCB]
{Center for Computational Biology, Flatiron Institute, New York, USA}
\author{David~Silva-Sánchez}
\affiliation[Yale University -Department of Mathematics]
{Department of Mathematics, Yale University, New Haven, CT, USA}
\alsoaffiliation[Flatiron Institute -CCM]
{Center for Computational Mathematics, Flatiron Institute, New York, USA}
\author{Julian~Giraldo-Barreto}
\affiliation[Flatiron Institute -CCB]
{Center for Computational Biology, Flatiron Institute, New York, USA}
\author{Bob~Carpenter}
\affiliation[Flatiron Institute -CCM]
{Center for Computational Mathematics, Flatiron Institute, New York, USA}
\author{Sonya~Hanson}
\affiliation[Flatiron Institute -CCM]
{Center for Computational Mathematics, Flatiron Institute, New York, USA}
\alsoaffiliation[Flatiron Institute -CCB]
{Center for Computational Biology, Flatiron Institute, New York, USA}
\author{Alex~H.~Barnett}
\affiliation[Flatiron Institute -CCM]
{Center for Computational Mathematics, Flatiron Institute, New York, USA}
\author{Erik~H.~Thiede}
\email{*ehthiede@flatironinstitute.org}
\affiliation[Flatiron Institute -CCM]
{Center for Computational Mathematics, Flatiron Institute, New York, USA}
\author{Pilar~Cossio}
\email{*pcossio@flatironinstitute.org}
\affiliation[Flatiron Institute -CCM]
{Center for Computational Mathematics, Flatiron Institute, New York, USA}
\alsoaffiliation[Flatiron Institute -CCB]
{Center for Computational Biology, Flatiron Institute, New York, USA}

\title{Ensemble reweighting using Cryo-EM particles}
\begin{document}

%\begin{tocentry}
%\begin{center}
% \includegraphics{TOC.png}
%\end{center}
%\end{tocentry}


\begin{abstract}
Cryo-electron microscopy (cryo-EM) has recently become a premier method for obtaining high-resolution structures of biological macromolecules. However, it is limited to biomolecular samples with low conformational heterogeneity, where all the conformations can be well-sampled at many projection angles.
While cryo-EM technically provides single-molecule data for heterogeneous molecules, most existing reconstruction tools cannot extract the full distribution of possible molecular configurations.
To overcome these limitations, we build on a prior Bayesian approach and develop an ensemble refinement framework that estimates the ensemble density from a set of cryo-EM particles by reweighting a prior ensemble of conformations, \textit{e.g.}, from molecular dynamics simulations or structure prediction tools.
Our work is a general approach to recovering the equilibrium probability density of the biomolecule directly in conformational space from single-molecule data.
To validate the framework, we study the extraction of state populations and free energies for a simple toy model and from synthetic cryo-EM images of a simulated protein that explores multiple folded and unfolded conformations.
\end{abstract}


\section{Introduction}

The last decade has seen a shift in the structural biology community from X-ray crystallography to cryo-electron microscopy (cryo-EM) for solving high-resolution structures of certain biological macromolecules. In cryo-EM, instead of using the diffraction pattern of a crystal of identical molecules, randomly-oriented 2D projection images of individual biomolecules at cryogenic temperatures are obtained with an electron microscope. In a typical cryo-EM dataset up to millions of noisy images or particles, are collected, and typically only a small percentage of the images that represents the most stable conformation is used to reconstruct the final density map. Recent improvements in cryo-EM due to the use of direct electron dectectors\cite{mcmullan2016direct}, motion correction\cite{li2013electron}, and fast reconstructions algorithms \cite{cossio2018likelihood} have led to a resolution revolution\cite{kuhlbrandt2014resolution} in cryo-EM, with many reconstructions now achieving atomic resolution\cite{Nakane2020}.

Concurrent with the resolution improvements in cryo-EM is a recent influx of methods for integrative-structural biology \cite{ward2013integrative,bottaro2018biophysical,bonomi2017principles} to understand biomolecular mechanisms by combining structural modeling and simulations with experimental data. These ensemble-refinement techniques, most notably maximum-entropy methods, extract the optimal weights of the ensemble members (\textit{e.g.}, conformations of the biomolecule from simulations) based on the data \cite{costa2022reweighting}. Typical experimental data that are used for ensemble refinement take the form of observables that are an ensemble average, such as nuclear magnetic resonance \cite{rieping2005inferential} or X-ray scattering \cite{rozycki2011saxs}. Many methodologies have been proposed as general approaches to reweight ensembles using these averaged observables \cite{boomsma2014combining,hummer2015bayesian, kofinger2019efficient,bonomi2016metainference,barrett2022simulation,roux2013statistical,cesari2018using}. 

Several integrative methods use a cryo-EM reconstruction as an averaged observable to refine structures. Current modeling tools fit or refine atomic coordinates into the map using molecular dynamics (MD) guided by the reconstructed map \cite{trabuco2009molecular,igaev2019automated,mori2021efficient,blau2022gentle,vuillemot2022nmmd}. Other approaches use the map to estimate the structural ensemble, for example, by determining the most probable set of conformations that construct an average map that is most correlated to the experimental map \cite{bonomi2018simultaneous}. However, despite cryo-EM giving access to millions of individual images, 
the reconstruction only contains information from the small percentage of particles that generated the high-resolution reconstruction (typically $<25$\%). Useful information about the system's conformational heterogeneity might be discarded along with the majority of particles discarded in the reconstruction process. 

Cryo-EM has the great advantage of being a single-molecule experiment. The cryo-EM freezing process is fast enough to trap the biomolecule in various conformations at near-native conditions \cite{Dubochet1988}. If one assumes that the freezing is instantaneous, then the conformations would be distributed according to the Boltzmann distribution at the temperature prior to freezing. Therefore, there is more information in the cryo-EM sample than just an average of the most probable state given by the 3D map. This has inspired the development of methods for extracting free-energy surfaces using individual particles \cite{Fischer2010,Dashti2014,Dashti2020} instead of 3D maps. Ensemble refinement has been proposed with the BioEM formalism \cite{cossio2013bayesian} to determine the minimal number of ensemble members that best represent the entire particle set (but without modifying their weights). More recently, the cryoBIFE method was developed to extract free-energy profiles along a predetermined 1D molecular path using a cryo-EM particle set \cite{giraldo2021bayesian}. The posterior distribution over free-energy profiles is then extracted within a Bayesian approach.  However, in cyroBIFE, the molecular path must be predetermined, which limits its applicability.

\begin{figure}[h!]
\centering  
\includegraphics[width=\columnwidth]{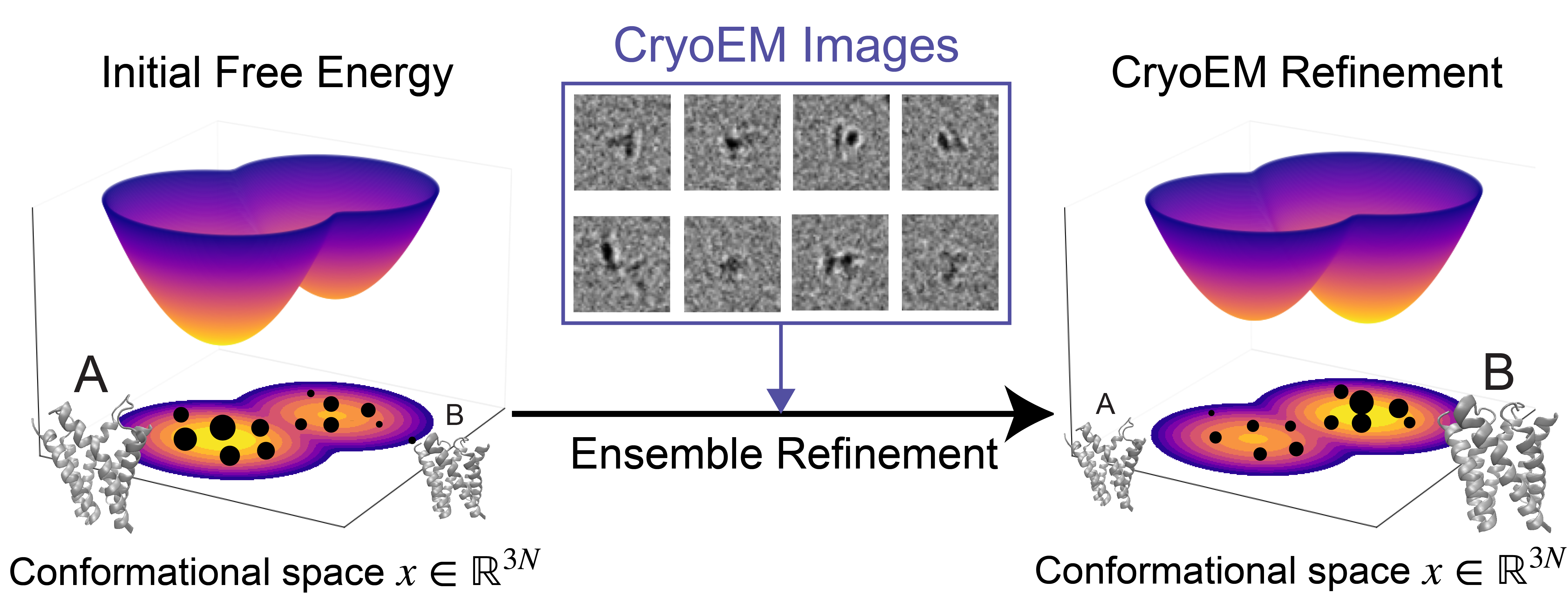}
\caption{
Ensemble reweighting using cryo-EM particles exemplified with a free-energy landscape. We take an initial guess for the system's conformational probability
density (left), provided by biomolecular simulations or modeling tools. 
We then correct the guess ensemble density
by sampling configurations $\{\cfg_t\}_{t=1}^T$ (black points) and comparing them to individual cryo-EM particles (middle) using a Bayesian framework.
The corrections are captured in a function $\rwt\!\left(\cfg;\weight\right)$ that depends on weights $\weight=\{\weight_t\}$.
The initial density guess multiplied by the correction $\rwt$ is an estimate for the probability density in conformational space (right). 
The example cartoon structure shown here is the semiSWEET transporter\cite{latorraca2017mechanism}.
}
\label{fig:fig1}
\end{figure}

In this work, we propose a single-particle cryo-EM ensemble reweighting method that builds on cryoBIFE. Instead of relying on the notion of a conformational path, our proposed method uses prior ensembles (from Rosetta\cite{das2008macromolecular,leman2020macromolecular} or MD), over which the ensemble density can be extracted using a posterior informed by the full set of cryo-EM particles (Figure \ref{fig:fig1}).
In the following, we present a general theory for ensemble refinement using statistically independent and identically distributed (i.i.d.)\ measurements (instead of averaged observables) within a Bayesian approach, building approximations of the ensemble density directly in conformational space. We first propose a simple approximation of the density assuming a uniform prior over the ensemble members and study the behavior of the recovered ensemble weights for a toy system (where the ``images'' are scattered data points) in different designed scenarios. We then propose a more general approximation for the ensemble density using clusters of sampled conformations from simulations or modeling tools. We apply this method to a prototypical benchmark multi-state peptide and demonstrate that this approach is able to retrieve the ensemble density underlying synthetic cryo-EM particles generated from conformations of MD simulations. We conclude with some future perspectives on the work.

\section{Theory}

\subsection{Biophysical background}

We assume that an experiment produces a set of samples of an observable measured
over individual conformations $\cfg$ of a biomolecule. In contrast to traditional reweighting methods that use averaged observables, in this case,
we treat experiments that give independent and identically distributed (i.i.d.)\ measurements. 
The biomolecule's
conformations live in a high-dimensional space of size three times the number of atoms of the biomolecule, $\cfg\in\mathbb{R}^{3\Natom}$.
The probability that the molecule adopts a configuration $\cfg$ is given by
the Boltzmann distribution,
whose probability density is given by
\begin{equation}
\truedensity(\cfg)=\frac{1}{Z_0} e^{-\beta H(\cfg)}~,
    \label{eq:Boltzmann}
\end{equation}
where $H$ is the molecular Hamiltonian, $Z_0=\int e^{-\beta H(\cfg)} \textrm{d}\cfg$ is the partition function, $\beta=1/k_BT$ with $k_B$ Boltzmann's constant, and $T$ the temperature.

If the Hamiltonian were known exactly, we would have full knowledge
of the biomolecule's conformational ensemble.
This information is valuable because one can not only calculate averaged observables but also extract free-energy surfaces $G$ along collective variables $s$ given by
\begin{equation}
    e^{-\beta G(s)}
        = \frac{1}{Z_0} \int \delta(S(x) -s)\, e^{-\beta H(x)} \textrm{d}x ~,
\end{equation}
where $S$ is the function that, given a configuration, returns the value of the collective variable.
One can use this to extract information about the metastable states of the system and their probability, as well as the activation barriers and transition states, which lead to the biomolecule's thermodynamic description.
Although we have some prior notions of biomolecular interactions, these are only rough approximations of the true Hamiltonian. For real-life biomolecules, the Hamiltonian $H$ is unknown. The goal of this work is to combine samples from an approximate $H$ with single-molecule (\textit{e.g.}, cryo-EM) data to recover a more accurate Boltzmann ensemble than the approximate $H$ would give alone.

\subsection{Ensemble reweighting for i.i.d.\ measurements}

We introduce a family of candidate probability densities over biomolecular configurations $p(\cfg\, |\weight)$, parameterized by a collection of parameters $\weight$.
Each parameterized density represents a possible hypothesis for the true Boltzmann ensemble.
We use Bayesian analysis to infer $\weight$ from the data  given by a collection of i.i.d.\ measurements $\ImgSet =\{\img_i\}$ with $i\in [1,\ldots,I]$ (for example, the image set in cryo-EM), where $I$ is the total number of data points.
Specifically, we seek to recover the \emph{posterior probability} density of $\weight$.
This density gives the probability that, conditioned on the observed data, $\weight$ specifies the true Boltzmann probability.
Recovering the posterior probability density (referred to as the posterior)
is the central objective of Bayesian algorithms.
With it, we can calculate the mean value of a predicted measurement,
as well as the corresponding uncertainty.
We discuss the precise quantities we estimate using the posterior in the Methods.

Bayes's theorem states that the posterior probability of the parameters is given by
\begin{equation}
    p\!\left( \weight  | \ImgSet  \right) \;\propto\; p\!\left(\weight\right) \, p\!\left( \ImgSet |  \weight \right)~, 
    \label{eq:bayes_rule}
\end{equation}
where $p\!\left(\weight\right)$ is the value of the prior distribution evaluated at $\weight$ (referred to as the prior) and
$p\!\left( \ImgSet |  \weight \right)$ is the likelihood of the measurements given a candidate probability density with parameters $\weight$.

In practice, the prior is typically given in closed form, and it can be evaluated easily.
To evaluate the likelihood, we assume that the biomolecules adopt their configuration independently of each other and that the data are generated independently through the same procedure.
Consequently, the likelihood of the data set is a product of the likelihood of the individual i.i.d.\ measurements
\begin{equation}
    p\!\left(  \ImgSet | \weight  \right) = \prod_{i} p\!\left(\img_i | \weight \right)~.
\end{equation}
Marginalizing over all configurations $\cfg$ that could have led to the $i$th observation gives $p(\img_i|\weight) = \int p(\img_i|\cfg)p(\cfg|\weight) \textrm{d}\cfg$, where $p\!\left( \cfg | \weight \right)$ gives the probability density of the molecule being in configuration $x$ given the parameter choice $\weight$.
The term $p\!\left(\img_i | \cfg\right)$ is the likelihood of observing $\img_i$ (\textit{e.g.}, an image) given that the biomolecule is in configuration $\cfg$; details for the toy model, and for cryo-EM likelihoods, are presented in the Methods below.
Thus, 
\begin{equation}
      p\!\left( \ImgSet | \weight \right) =  \prod_{i}  \left( \int p\!\left(\img_i | \cfg \right)  p\!\left( \cfg | \weight \right) \textrm{d} \cfg  \right).
    \label{eq:marginal_over_configuration}
\end{equation}
Note that here the dummy variable $x$ is local to each of the integrals in the product.

The variable $\cfg$ takes values in a very high-dimensional space ($3 \Natom$ degrees of freedom). Therefore,
the integrals in Eq.~\ref{eq:marginal_over_configuration} are generally intractable. This makes the construction of flexible and physically realistic probability densities a daunting task.
Fortunately,
we can leverage prior work on building ensembles for biomolecular systems.
Rather than building our candidate densities from scratch,
we instead take an existing guess for the density and apply a multiplicative reweighting factor
to adjust for errors in the guess.
To formalize this refinement procedure, we write the family of candidate densities as the product
\begin{equation}
    p\!\left( \cfg | \weight \right) =
            \rwt\!\left(\cfg;\weight\right)
            \pguess\!\left( \cfg \right)~,
    \label{eq:reweighted_density}
\end{equation}
where $\pguess$ is the initial guess for the Boltzmann density that we seek to correct, and $\rwt$ is a multiplicative correction that depends on parameters $\weight$.
By varying $\weight$ we can tune the correction to bring $p\!\left(\cfg | \weight \right)$ closer to the system's ensemble measured in the data.
Note that we explicitly require $p(\cfg | \weight)$ to be a valid probability density.
Consequently, $\rwt$ must be nonnegative everywhere and must be scaled such that $p\!\left(\cfg| \weight \right)$ integrates to 1.

Substituting the candidate density from Eq.~\ref{eq:reweighted_density} into Eq.~\ref{eq:marginal_over_configuration} gives the following expression for the likelihood
\begin{equation}
    p\!\left(\ImgSet | \weight \right) =
    \prod_{i} \left( \int
            p\!\left(\img_i | \cfg \right)
            \rwt\!\left(\cfg;\weight\right)
            \pguess\!\left( \cfg \right)
        \textrm{d} \cfg  \right)~.
    \label{eq:raw_likelihood_with_correction}
\end{equation}
Substituting this expression into Eq.~\ref{eq:bayes_rule} enables sampling the posterior, for example, with Markov chain Monte Carlo (MCMC) methods to extract estimates of $\weight$ given the data (see the Methods section for details).
However, the computational cost of evaluating the integrals in Eq.~\ref{eq:raw_likelihood_with_correction} is large. Moreover, the accuracy of the estimates will depend strongly on the choice of $\pguess$ and of $\eta$. In cryoBIFE \cite{giraldo2021bayesian} an approximated probability density was proposed that relied on a predetermined 1D conformational path. To work directly in conformational space and overcome these issues, in the following, we present two approaches of approximating $p\!\left( x | \weight \right)$ of increasing complexity. 

\subsubsection{Simple ensemble reweighting}\label{ER-simple}

To connect with traditional ensemble reweighting methods \cite{costa2022reweighting}, we first approximate the parameterized density by a discrete set of conformations $\cfg_t$ for $t = 1,\dots,T$, where the total number of ensemble members $T$ is much smaller than $I$, the number of data points.
Each conformation $\cfg_t$ is a point in $\mathbb{R}^{3\Natom}$.
We assume that the set is a sufficiently good representation of the whole conformational space.
A simple initial guess for the Boltzmann density is to be uniform over these conformations,
\begin{equation}
    \pguess(x) = \frac{1}{T} \sum_{t} \delta\left(\cfg - \cfg_t\right)~.
    \label{p0deltas}
\end{equation}
Multiplying this initial density by any choice of $\eta$ reduces to scaling each delta function individually.
Hence we define the set of free parameters $\weight=\left\{\weight_t\right\}$ to be the magnitude of these scalings given to the candidate density
\begin{equation}
    p\!\left( x | \weight \right) = \sum_{t} \delta( \cfg - \cfg_t) \weight_t~,
    \label{eq:simple_reweighting_density}
\end{equation}
where we have incorporated the factor of $1 / T$ into the scalings for simplicity.
To ensure that Eq.~\ref{eq:simple_reweighting_density} gives a valid probability density,
we require that  each weight be non-negative and that $\sum_t \alpha_t=1$.
Using Eq.~\ref{eq:simple_reweighting_density} as the model probability density, the posterior obeys
\begin{equation}
    p\!\left( \weight | \ImgSet \right) \;\propto\;
    p\!\left(\weight\right) \prod_{i} \left( \sum_t p\!\left( \img_i | \cfg_t\right) \weight_t \right)~.
    \label{eq:post_weights_simple}
\end{equation}
In contrast to  traditional ensemble reweighting methods that use averaged observables \cite{costa2022reweighting}, this posterior expression takes into account the individual observations. However, in both cases, the conformational ensemble only consists of the given $\{\cfg_t\}$, and the ensemble weights are to be extracted. We use MCMC sampling for the latter. However, maximum \textit{a posteriori} methods\cite{hummer2015bayesian} could also be used. We note that $T$ is computationally limited to small values with this approach.

\subsubsection{Approximating the ensemble density on clusters of conformational samples}\label{ER-cluster}

For realistic scenarios, having a small and \textit{a priori} chosen set of configurations might not be sufficient to give a good representation of the system's full Boltzmann ensemble.
Fortunately, the computational biophysics and biochemistry community have put extensive effort into building computational models for approximating the Boltzmann distribution of biomolecular systems,
as well as developing algorithms that can generate sample configurations from these models.
For example, MD simulations use explicit models of the system's Hamiltonian known as \emph{force fields}\cite{FrenkelSmit2002},
and generate samples from the system's Boltzmann distribution.
Packages intended to find the folded structure of proteins, such as Rosetta\cite{das2008macromolecular,leman2020macromolecular}, often use a similar approach,
employing heuristic force fields and computationally efficient algorithms to guide the generation of new configurations.
More recently, machine-learning approaches to generating molecular ensembles have seen some success.
Deep probabilistic models\cite{noe2019boltzmann}
can directly generate candidate protein conformations.
Structures generated by neural networks can also be combined with MD to enrich and refine the structural ensemble\cite{degiacomi2019coupling,vani2022sequence}.
This diversity of approaches provides a rich pool of possible choices to use for $\pguess$.

In all of these cases, $\pguess$ is typically too complex to allow us to evaluate the integral in Eq.~\ref{eq:raw_likelihood_with_correction} explicitly.
However, we can approximate the integral using sample averages.
As above, we assume that we have access to a collection of $\Npoints$ configurations, $\left\{x_t\right\}$ with ${t=1,\ldots,\Npoints}$,
drawn from $\pguess$.
However, we now use the more general reweighting approach with Eq.~\ref{eq:reweighted_density} that will be controlled by a parameter vector $\alpha$ with much less than $T$ components.
Therefore, approximating the integral in Eq.~\ref{eq:raw_likelihood_with_correction} over samples, gives an estimate of the likelihood of the data set
\begin{equation}
    \bar{p}\!\left( \ImgSet | \weight \right) \;=\; \prod_i
    \left(
    \frac{1}{\Npoints}\sum_{t} p\!\left(\img_i | x_t \right)
            \rwt\!\left(x_t;\weight\right) \right)~,
    \label{eq:empirical_likelihood}
\end{equation}
where we have used the overbar ($\bar{\, \cdot \,}$) to indicate that this is an average over sampled configurations.
Similarly, we approximate the normalization constraint
by requiring that $\frac{1}{\Npoints} \sum_{t} \rwt\!\left(x_t;\weight\right)=1$.
Note that evaluating both Eq.~\ref{eq:empirical_likelihood} and the approximate normalization constraint
does not require an explicit functional form for $\pguess$ but only samples from it.

$\pguess$ is likely to capture the atomistic and chemical features of the biomolecule correctly. For example,
MD force fields give accurate bond lengths and intramolecular angles.
However, they have more difficulty with capturing more large-scale, collective properties.
In particular, the relative probabilities of various free-energy minima (\textit{i.e.}, metastable states)
are likely to be incorrect. A well-chosen $\rwt$ should be capable of correcting these errors.
Consequently, we will build the reweighting function by dividing the
configuration space into $M$ disjoint clusters that we define by applying a clustering algorithm to $\left\{ x_t \right\}$.
We define $\rwt$ piecewise on each cluster $m$, where the parameters $\weight=\{\weight_m\}$
are the magnitude of $\eta$ on each.
Let $\1_m$ be the indicator function on each cluster, \textit{i.e.},
the function that is $1$ for configurations belonging to cluster $m$ and $0$ for all others,
then we write $\rwt$ as
\begin{equation}
    \rwt\!\left(x ; \weight \right) = \sum_{m} \weight_m \, 
    \1_m\!\left(x\right)~,
\end{equation}
with $m=1,\ldots,M$.
By substituting this expression into Eq.~\ref{eq:empirical_likelihood} the likelihood estimate becomes
\begin{equation}
\bar{p}\!\left(\ImgSet | \weight  \right) =  \prod_i \left(
 \frac{1}{\Npoints} \sum_{t}  \sum_{m} p\!\left(\img_i | x_t \right)\weight_m \, \1_m\!\left(x_t\right) \right)~.
    \label{eq:ll_with_samples}
\end{equation}
Note that the constraint that $p\!\left(\, \cfg\, ; \alpha \right)$ be a valid probability density
requires that $\alpha$ have non-negative entries. Moreover,
if $N_m$ is the number of conformations that are in the $m$'th cluster,
we can reduce our constraints on $\weight$ to the requirement $\frac{1}{\Npoints}\sum_m N_m \weight_m = 1$.

When the number of images and sampled configurations becomes large, evaluating
$p\!\left(\img_i | x_t\right)$ for every possible datapoint-configuration pair can be computationally expensive (\textit{e.g.}, for cryo-EM).
Therefore, we make the additional simplifying assumption that the clusters are spatially compact, where
 we expect $p\!\left(\img_i | \cfg \right)$ to vary little within the cluster.
We then approximate the likelihood by the value at the medoid of the cluster $\ccenter_m$,
\begin{equation}
    \bar{p}\left( \ImgSet | \weight \right)
    \approx  \prod_i \left(\frac{1}{T}
        \sum_{m} p\!\left(\img_i | \ccenter_m \right)\weight_m \, N_m \,
             \right)  ~.
    \label{eq:final_ll_on_clusters}
\end{equation}
Substituting this approximation into Eq.~\ref{eq:bayes_rule} gives, up to a multiplicative factor,
an approximation for the posterior probability of the parameterized density
given the observations,
\begin{equation}
    p\!\left( \weight |  \ImgSet\right)
    \propto p(\weight) \prod_i \left(\frac{1}{T}
        \sum_{m} p\!\left(\img_i | \ccenter_m \right) \weight_m \, N_m \,
             \right) ~.
    \label{eq:final_post_on_clusters}
\end{equation}
This can be directly used in
MCMC to estimate the expected $\weight$, which entails the approximation of the ensemble density. An advantage of having the ensemble density represented in the full molecular space $\mathbb{R}^{3\Natom}$ is that one can calculate free-energy landscapes over collective variables chosen \textit{a posteriori}, as a post-processing step (see the Methods below).

\section{Methods}\label{sec:methods}

\subsection{Markov chain Monte Carlo}

We used an MCMC method to sample the ensemble weights $\{\weight_j\}$ with $j\in \{ 1,\ldots,J \}$ from the posterior defined in Eq.~\ref{eq:post_weights_simple} with $J=\Npoints$ and Eq.~\ref{eq:final_post_on_clusters} with $J=M$ for the simple reweighting and clustering reweighting aproaches (sections \ref{ER-simple} and \ref{ER-cluster}), respectively. We initialized the values of 8 MCMC chains by drawing from a Dirichlet distribution of dimension $J$. The MCMC chains are sampled using Hamiltonian Monte Carlo (HMC) with the no-U-turn sampler (NUTS) \cite{hoffman2014nuts}. Each of the 8 chains undergoes 1000 warmup steps, then 10,000 sampling steps to generate a sample of 80,000 draws of $\weight$. The MCMC algorithm is implemented with Stan \cite{carpenter2017stan}. Convergence diagnostics, such as $\widehat{R}$ and effective sample size (ESS), for the MCMC are described in the Supplementary Text.

\subsection{Toy model data and likelihood}

To study the simple reweighting approach (section \ref{ER-simple}), we use a toy model with normally distributed data. A data point, which can be thought of as an ``image'' in the abstract sense, is a positional vector $\img_i \in \mathbb{R}^P$ derived from a ``conformation'' $x \in \mathbb{R}^N$. For simplicity, we choose $P=N=3$. The data points $Y=\{\img_i\}$ are drawn from three normal distributions, having the same scale (\textit{i.e}, standard deviation) $\lambda$, and located at centers $\cfg_A$, $\cfg_B$, and $\cfg_C$. Each distribution has a different weight, with $\weight_A=0.5$, $\weight_B=0.3$, and $\weight_C=0.2$, and the weights sum to one. The scale $\lambda$ emulates the noise in images, which is set to one considered known. In total, 10,000 data points are generated. These are shown as a scatter plot in Figure~\ref{fig:fig2}.

Given the toy model data, we apply the simple reweighting approach to infer the weights for different sets of $\{\cfg_t\}$, ranging from the true distribution centers to misplaced $\cfg_t$ or sets with a higher number of members than the true centers (Table \ref{tab:toy_model_optweights}). We use Eq.~\ref{eq:post_weights_simple} with likelihood
$p_{\mathrm{toy}}(\img_i|\cfg_t) = (2\pi\lambda^2)^{-3/2} \, \exp\!\left(-\frac{||\img_i-\cfg_t||_2^2}{2\lambda^2}\right)~$,
where $||.||_2=$ is the $\ell^2$-norm that measures the distance between points $\img_i$ and $\cfg_t$ in $\mathbb{R}^3$. For simplicity, we fix $\lambda$ to its known value rather than inferring it. We use the MCMC sampling method described above to extract the expected weights $\{\weight_t\}$ for the different $\{\cfg_t\}$ sets.

\subsection{Cryo-EM imaging model}

Following ideas from previous literature
\cite{seitz2019simulation,bendory2020single,grant2010image}, we use a simple model to represent the image-formation process in cryo-EM. Starting from a molecular configuration $x$, we assume that the electron density $\rho(x)$ is the sum of spherically-symmetric 3D normal densities centered at the $x$ atom positions, all of which are assumed to have the same scale. We use a weak phase approximation to model the projection image. The forward model consists of a rotation $R_\phi$ of $\rho(x)$, then a projection $P_z$ along the $z$ axis, a point-spread function convolution $\mathrm{PSF}_\theta$ (equivalent to using the contrast transfer function (CTF) in Fourier space) with parameters $\theta$ that includes the defocus, and a translation. For simplicity, we set the biomolecule's center of mass at the image center so that there is no uncertainty in the particle center. This simple imaging model is
\begin{equation}
\mathrm{Img}_{\theta,\phi}(x)=\mathrm{PSF}_\theta \, \mathcal{P}_z \,\mathcal{R}_\phi\, \rho(x) ~.
\label{eq:forward_cryo}
\end{equation}
In practice, $\mathrm{Img}_{\theta,\phi}(x)$ is discretized as a 2D grid with number of pixels $N_{\mathrm{pix}}$. Additional details of the imaging model are provided in the Supplementary Text.
We use this forward model in the cryo-EM likelihood to compare a configuration with an image and to generate synthetic images by adding normal white noise (details below).

\subsection{Cryo-EM likelihood}

A crucial part of Eq.~\ref{eq:raw_likelihood_with_correction} is $p\!\left( y_i | x \right)$ the likelihood that compares a single observation to a configuration. In the case of cryo-EM, $y_i$ is an individual particle image of size $N_{\mathrm{pix}}$. We assume that the measured image is a noisy representation of the forward model $\mathrm{Img}_{\theta,\phi}$ from configuration $x$ (Eq.~\ref{eq:forward_cryo}).  We assume i.i.d.\ normal white noise with scale $\lambda$ at each pixel. The likelihood of this noise model is
\begin{equation}
    p\!\left( y_i | x \right) = (2\pi\lambda^2)^{-N_{\mathrm{pix}}/2} \exp\!\left(-\frac{||y_i -\mathrm{Img}_{\theta,\phi}(x)||_2^2}{2\lambda^2}\right)~,
    \label{eq:likeli_cryo}
\end{equation}
where $||\cdot||_2^2$ denotes the squared $\ell^2$-norm, \textit{i.e.}, $||a-b||_2^2=\sum_{l}(a_l-b_l)^2$ with $l \in \{1,\ldots,\Npix \}$.
For simplicity, we assume that we know the parameters $\phi$, $\theta$ and $\lambda$, \textit{i.e.}, the optimal pose of the experimental image, the CTF parameters, and the colorless noise standard deviation. We note that this is a large simplification of the problem, as a major challenge in cryo-EM is finding the optimal projection direction. Nonetheless, this simplification does not undermine the reweighting theory developed here, and computational approaches exist to fully evaluate the image-conformation likelihood \cite{cossio2013bayesian,cossio2017bioem,Cossio:Micro:2018}.

\subsection{Structure-generating molecular dynamics}

An unbiased MD simulation of Chignolin is performed using the Amber ff99SB-ILDN force field, \cite{lindorff2010improved} with GROMACS 2022.1 \cite{abraham2015gromacs}. The initial structure of Chignolin is an experimentally resolved structure obtained from the Protein Data Bank (PDB ID: 1UAO)\cite{honda2004chignolin}. The PDB structure is solvated in explicit water with TIP4P water model \cite{lawrence2003tip4p} in a cubic box with 1 nm buffer to all periodic boundaries from any protein atom to avoid non-bonded interactions of the protein with its periodic image. Two Na$^+$ ions are added to the box to neutralize the electrostatic charges of the protein. The system underwent energy minimization steps until all interatomic forces are less than 1000 kJ/mol/nm to resolve steric conflicts. Both equilibrating and data-producing MD steps use the leapfrog integrator with 2-fs step size and the Berendsen thermostat was used with $\tau_t=0.1$ to maintain the temperature of the system at 300 K. For non-bonded interactions, the Verlet cutoff scheme is used on both VDW and Coulomb forces with cutoff distances of 1.0 nm for both cases. The system is first equilibrated in the NVT ensemble for 100 ps with bond constraining LINCS algorithm \cite{hess1997lincs}, then in NPT ensemble for another 100 ps with Parrinello-Rahman pressure coupling. The NPT equilibrated system is then duplicated into 6 replicas with 6 different random seeds for initial velocity generation. After the velocities are generated, the 6 replicas are further equilibrated for 100 ps, then run for 2 $\mu$s of unbiased MD simulation, generating 12 $\mu$s of MD trajectory (120,000 frames) in total that serves as the structure-generating trajectory for this study. 

\subsection{3D Clustering}

Even with knowledge of the optimal pose, computing the pairwise distance between hundreds of thousands of structures to hundreds of thousands of images is too computationally costly. To reduce this, we cluster the conformations from the MD trajectory using k-medoids clustering \cite{schubert2022kmedoids,schubert2021fast}. This clustering method requires as input the number of clusters $M$ and a distance matrix between the objects to be clustered, which we define as the C$_\alpha$ RMSD between every pair of conformations. The algorithm returns the cluster centers $\{\ccenter_m\}$, which provide a representative subset of the 120,000 MD conformations. To assess the results for different numbers of clusters, we use the $\Ncenclusters \in [10,20, 50, 100]$. 
For each $M$, the cluster centers $\{\ccenter_m\}$ are used as input in the reweighting scheme with Eq.~\ref{eq:final_post_on_clusters}  to obtain the weight $\weight_m$ associated with each $\ccenter_m$. In the Supplementary Text, we present an alternative clustering method \cite{klem2022size} that we used to compare the results.  

\subsection{Metastable state classification}

Each MD conformation is assigned into a folded, unfolded, and misfolded metastable state. This was done by clustering the MD conformations using the algorithm described above with $M=3$ that results in a cluster center medoid  for each metastable state (see Supplementary Figure S1).
Structures with C$_\alpha$ RMSD less than 1.2 \AA\ from the folded medoid, or misfolded medoid, are classified into the folded or misfolded state, respectively. All other conformations that do not fall into these categories are considered unfolded. 

\subsection{Image-generating ensemble}

For the purpose of generating synthetic cryo-EM images of structures from a different ensemble distribution, an MD trajectory of Chignolin is used from D.E. Shaw Research, \cite{kresten2011how} which was generated using a different force field CHARMM22* \cite{piana2011robust}. It is a 106 $\mu$s long simulation trajectory, with 106,949 frames, that serves as the image-generating trajectory for this study. We generate one image per frame, resulting in 106,949 synthetic images in total. We rotate each image using an independent rotation matrix drawn uniformly from $\textrm{SO}(3)$ before projecting with $P_z$. 
For the CTF parameters, the defocus is drawn uniformly between 0.027 and 0.090 \AA$^{-1}$, and the b-factor and amplitude are drawn from a uniform distribution $\in [0,1]$. 
Then, we apply the imaging model from Eq.~\ref{eq:forward_cryo}. The image size is $\Npix=256\times256$, and the pixel size is 0.15 \AA.
Then, i.i.d.\ normal white noise is added to every pixel with $\lambda$ defined such to have different signal-to-noise ratios, SNR $\in$ [1,$10^{-1}$,$10^{-2}$,$10^{-3}$,$10^{-4}$]. The power of the signal of an image P$_\mathrm{signal}$ is the mean squared intensity of the pixels within a circular area of radius = 0.4$\times$image width. Normal white noise of scale $\lambda$ is added to the image, with the scale being calculated using the SNR, where $\textrm{SNR} = \textrm{P}_\mathrm{signal}/\lambda^2$. Example images at different SNRs are shown in Supplementary Figure S2. Ground truth populations of this trajectory are determined by summing the number of conformations belonging to each state that were classified using the procedure described above.

\subsection{Free-energy landscape calculation}

We can generate interpretable and physically-meaningful descriptors such as free-energy surface $G$, as defined in the Theory, with the formalism described in section \ref{ER-cluster} based on approximating the ensemble density using cluster from conformational samples. Since the method determines the density in the $\mathbb{R}^{3\Natom}$ atomic-coordinate space, we can potentially choose any set of collective variables to generate a free-energy landscape during post-processing.
In general, for any given value of $\weight$, we can estimate observable averages using the estimated density, $p\!\left(\,\cfg\,| \weight\right)$. This corresponds to evaluating a sample mean over our collection of configurations. 
Denoting our observable as $f$, we estimate its average as
\begin{equation}
    \int f\!\left( \cfg \right) p\!\left(\cfg | \weight \right) \textrm{d} \cfg 
        \approx \frac{1}{T} \sum_t \sum_m f\left(x_t\right) \weight_m \,
        \1_m\!\left(x_t\right) ~,
\end{equation}
where we have used $p\!\left(\cfg | \weight \right) =\rwt\!\left(x ; \weight \right)\pguess(\cfg)$ described in section \ref{ER-cluster}. Examining the distribution of ensemble average estimates over values of $\weight$ allows us to construct posterior mean estimates and credible intervals, for example, of a free-energy landscape $G$ over collective variables $s$.
In practice, to reconstruct the free-energy surface, the cluster-center weights are allocated to all the MD frames (from the structure-generating trajectory) by assigning $\weight_m$ of the cluster centroid to each of the conformations in cluster $m$, \textit{i.e.}, $\weight_t=\weight_m$ for $x_t$ in $m$. The two collective variables used for the 2D free-energy surface of Chignolin are the C$_\alpha$ RMSD with respect to the reference folded (RMSD$_\text{folded}$) and to the misfolded (RMSD$_\text{misfolded}$) cluster centers found by k-medoids clustering (described above).

\section{Results}

\subsection{Simple ensemble reweighting for a toy model}

\begin{figure}[h!]
\centering  
\includegraphics[width=0.8\columnwidth]{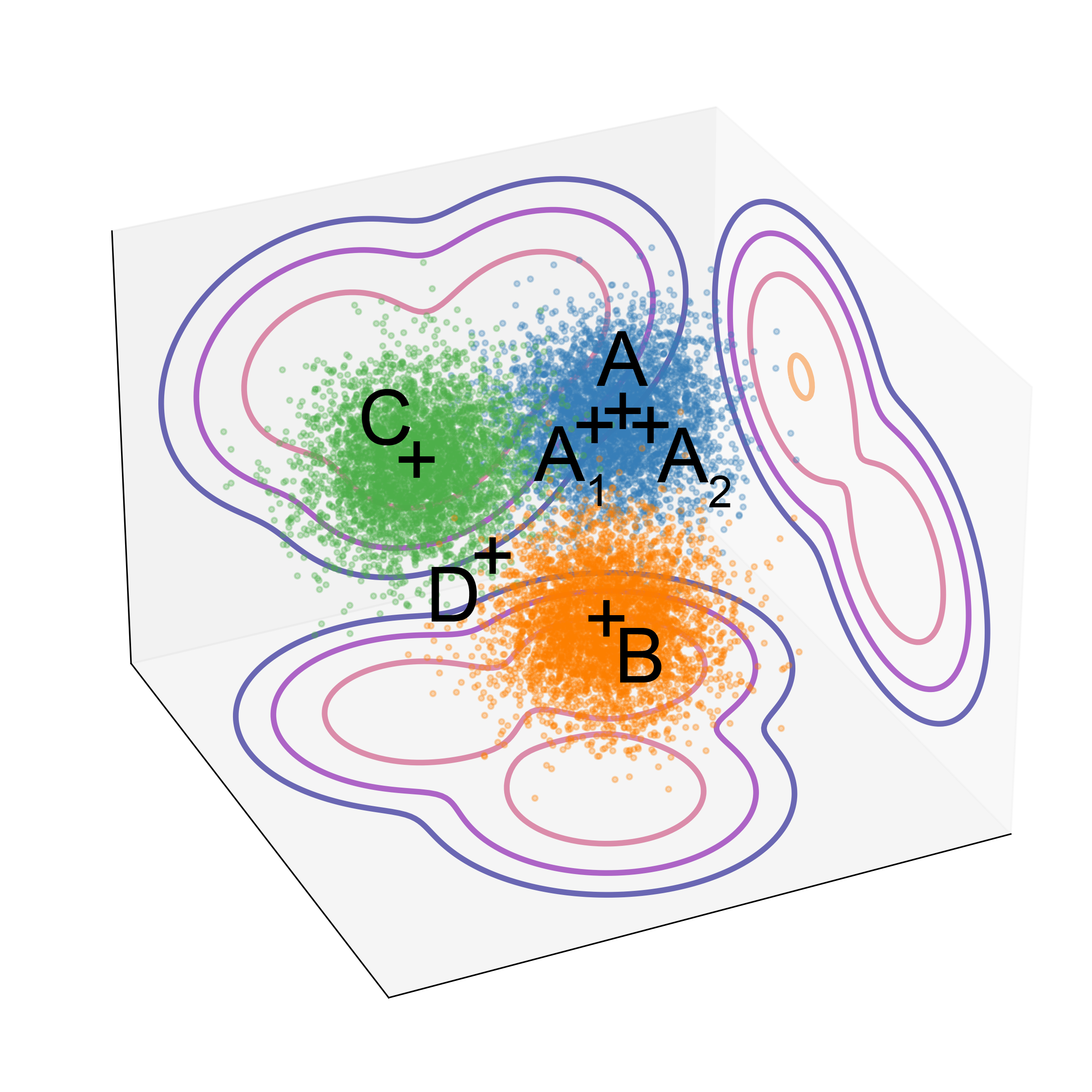}
\caption{Toy model with 3D normally distributed data. The blue, orange, and green points are generated from three separate normal distributions with proportions 0.5, 0.3, and 0.2, respectively, representing three metastable states of a system. The contour maps on the $x$-$y$, $y$-$z$, and $x$-$z$ planes are the projected free-energy surface along $z$-, $x$-, $y$-axis, respectively. The crosses with annotated letters \{A, A$_1$, A$_2$, B, C, D\} indicate the position of the ensemble members $\{\cfg_t\}$ placed to extract the weights $\{\weight_t\}$ shown in Table \ref{tab:toy_model_optweights}.
}
\label{fig:fig2}
\end{figure}

In this section, we use a simple toy model to study how well the simple ensemble reweighting approach (section \ref{ER-simple}), which uses a density approximation with a uniform prior, recovers state populations, even in cases where the ensemble members are not ideal or where there are multiple members within a state.   
For this toy model, the data are drawn from three separate normal distributions (henceforth metastable states) with different populations (Figure \ref{fig:fig2}). For this example, the data and ensemble members live in the same space $\mathbb{R}^3$, and the likelihood between a data point and ensemble member $\{x_t\}$ is presented in the Methods.

We explore the behavior of the optimized weights $\{ \weight_t\}$ for different $\{\cfg_t\}$ sets using the data presented in Figure \ref{fig:fig2}. The first case (\textit{i}) is the ideal scenario where three $\cfg_t$ are placed at the centers of the normal distributions used to generate the data at positions A, B, and C. This emulates perfectly placed $\{\cfg_t\}$ to sample the weights. We applied the simple ensemble reweighting approach, assuming a uniform prior and a small set of ensemble members (section \ref{ER-simple}). We compare the expected weights to the true populations (Table \ref{tab:toy_model_optweights}). As expected, the weights and state populations are in good agreement.
For the second case (\textit{ii}), we add a new ensemble member point D in Figure~\ref{fig:fig2}, which is the mid-point between centers B and C. This scenario emulates having an underrepresented ensemble member and having more points $\{\cfg_t\}$ than the intrinsic metastable states. Adding an extra member D in low-dense region results in D having close-to-zero weight, and the relative density of the three other members (A, B, and C) remains unchanged (Table \ref{tab:toy_model_optweights}). This implies that if the ensemble has members that are not representative of the data (\textit{e.g.}, unpopulated conformations), our method assigns negligible weights to them. This is an advantage because a ``perfect'' ensemble (with members at the state centers) is not required, and underrepresented members do not hinder the results. In the third case (\textit{iii}), we place four $\{\cfg_t\}$ at points A$_1$, A$_2$, B, and C (Figure \ref{fig:fig2}), emulating a cluster being sampled by two proximal members instead of one perfectly placed center A. With a split metastable state A$_1$ and A$_2$, the weights of B and C remain unchanged, while A$_1$ and A$_2$ split almost in half the population of the state generated from center A. This shows that when there is more than one member in the same state, the weights of that cluster will be split, and the relative weights within the state will depend on the local density (similarly to case (\textit{ii})). These results show that the algorithm is able to retrieve the weights $\weight$ for ensembles with different representative members, which do not necessarily have to be placed on the centers of the metastable states.

\begin{table}[h!]
\centering
\scalebox{0.8}{
\begin{tabular}{|l|cccccc|}
\hline
 $\{\cfg_t\}$         & A       & A$_1$            & A$_2$            & B        & D        & C        \\
\hline
True Population & 0.5     &                &                & 0.3      &          & 0.2      \\
\hline
Case (\textit{i})  & 0.501 $\pm$ 0.004 & --- & --- & 0.300 $\pm$ 0.004 & --- & 0.200 $\pm$ 0.003 \\
Case (\textit{ii})  & 0.501 $\pm$ 0.004 & --- & --- & 0.300 $\pm$ 0.003 & 0.001 $\pm$ 0.001 & 0.199 $\pm$ 0.003 \\
Case (\textit{iii})  & --- & 0.26 $\pm$ 0.03 & 0.24 $\pm$ 0.03 & 0.300 $\pm$ 0.004 & --- & 0.199 $\pm$ 0.003 \\
\hline
\end{tabular}
}
\caption{Toy model with 3D normally distributed data (shown in Figure~\ref{fig:fig2}). The true cluster population is related to the relative number of data points drawn from the normal distributions, which are centered at positions A, B, and C. The expected weights $\{\weight_t\}$ are shown for cases (\textit{i}), (\textit{ii}) and (\textit{iii}) with different ensemble members $\{\cfg_t\}$. The estimated uncertainty is shown as the standard deviation in MCMC samples. %
}
\label{tab:toy_model_optweights}
\end{table}

\subsection{Cryo-EM ensemble reweighting using sample conformations}

\begin{figure}[h!]
\centering
\includegraphics[width=0.75\columnwidth]{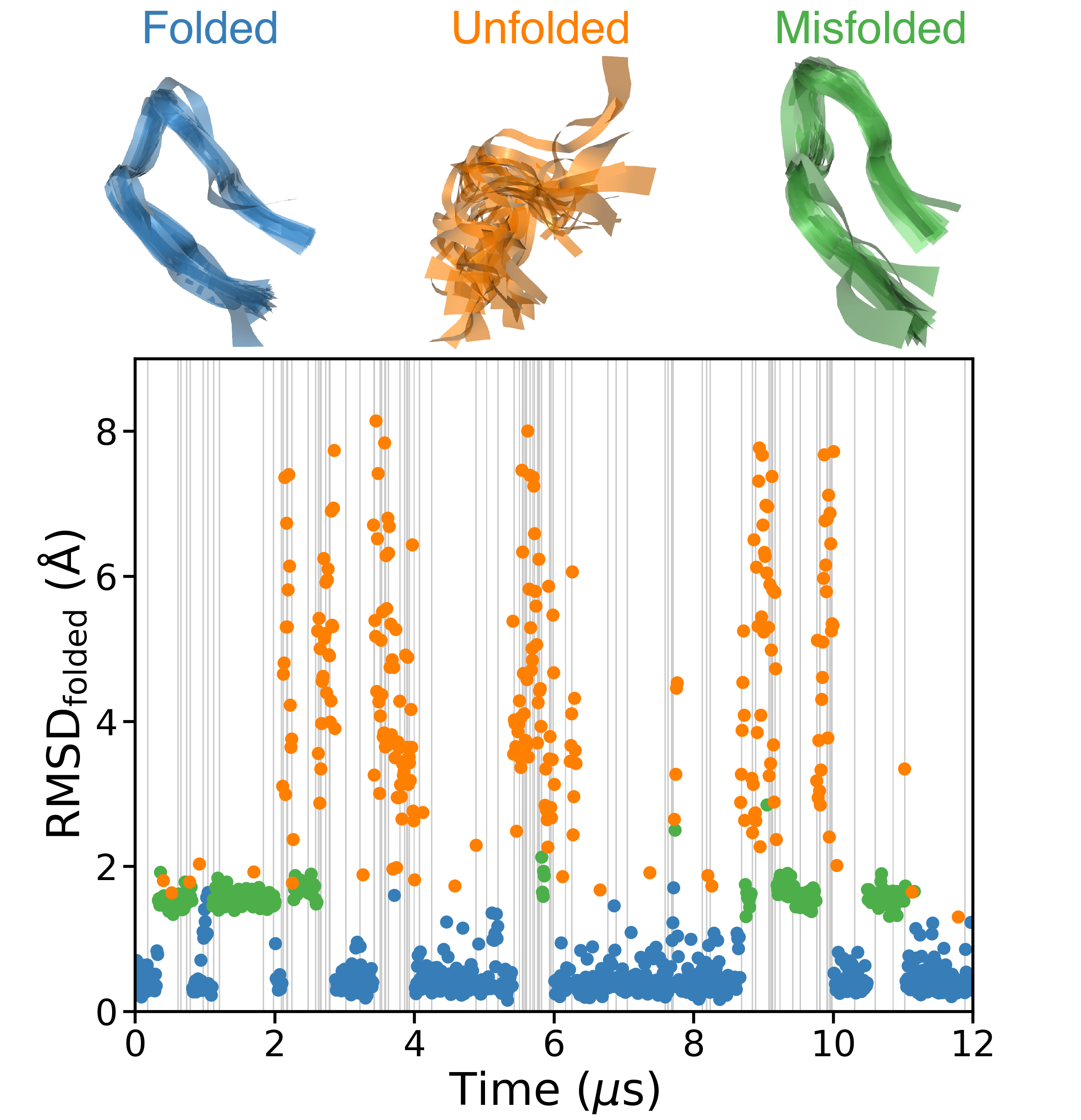}
\caption{C$_\alpha$ Root-mean-squared deviation (RMSD) 
of the structure-generating MD trajectory with respect to the folded state, RMSD$_\text{folded}$,
as a function of simulation time. The simulation time is across replicas: [0,2] $\mu$s indicates the first replica, [2,4] $\mu$s indicates the second replica, and so on. RMSD$_\text{folded}$ is calculated with respect to the reference folded structure. 1200 frames are shown as circles. Blue indicates members of the folded state. Green indicates members of the misfolded state. Orange  indicates members of the unfolded state. Grey bars indicate the position of the $M=100$ k-medoids cluster centers (see the Methods). 20 conformation from each metastable state is shown with a ribbon representation on top.
}
\label{fig:fig3}
\end{figure}

We performed MD simulations to sample conformations of the peptide Chignolin (see the Methods for details). Chignolin adopts three major metastable states: ``folded'', ``misfolded'', and ``unfolded'' states \cite{satoh2006folding}, visualized in Figure~\ref{fig:fig3}-top. The folded state is the most probable state, which consists of a single anti-parallel $\beta$-hairpin with aligned termini. The misfolded state is a non-native metastable state, which consists of a $\beta$-hairpin with the termini offset by one residue. In this study, the folded and misfolded states serve as two structurally similar metastable states with just small conformational differences. The unfolded state is defined by the conformations that neither resemble the folded nor the misfolded state (see the Methods).

We obtained conformations of Chignolin from two independent MD simulations: 6 replicas of 2 $\mu$s (12 $\mu$s in total) of MD simulations (structure-generating trajectory, Figure \ref{fig:fig3}), and 106 $\mu$s MD simulations from DESRES \cite{kresten2011how} (image-generating trajectory). These simulations were performed with different force fields and conditions, and they have different populations for the metastable states, which serve as a benchmark to test the effectiveness of the ensemble reweighting.
The 120,000 conformations from the structure-generating trajectory are clustered by k-medoids using the C$_\alpha$ RMSD as distance (see the Methods). The cluster centers compose the $\{\ccenter_m\}_{m=1}^M$ and different numbers of clusters $M$ (10, 20, 50, and 100) are used to examine the effect of having different members for the reweighting. In Supplementary Table 1, we show the different numbers of cluster centers belonging to each metastable state.

Conformations from the image-generating trajectory are used to generate synthetic cryo-EM images. Each MD conformation from this trajectory is used to generate the synthetic images (see the Methods). To assess the performance of the algorithm at different noise levels, normal white noise is added at $\textrm{SNR} \in [10^0, 10^{-1}, \ldots, 10^{-4}]$. For comparison, we also generate an image set with no noise. We first focus on the $\{\ccenter_m\}$ that has 20 cluster centers with 4, 13, and 3 centers belonging to the folded, misfolded, and unfolded states, respectively. The weight for each center is retrieved using the reweighting methodology defined in section \ref{ER-cluster} that approximates the ensemble density on clusters from conformational samples. The population of each state is calculated by summing the weights of the cluster centers belonging to each metastable state. We show the retrieved state population in Table~\ref{tab:replica_desres_optweights} for image sets with different SNRs.
At high SNR (SNR $\geq 10^{-2}$), the algorithm is able to retrieve the relative population of the three states with good accuracy. This demonstrates that the reweighting algorithm is able to refine the MD ensemble using the information from noisy cryo-EM images, which are given by an independent and different ensemble distribution.
Assuming well-defined poses, our algorithm is able to retrieve the ensemble density until SNR $=10^{-3}$. 
However, as was shown for cryoBIFE \cite{giraldo2021bayesian}, we expect the recovery to fail for higher SNR if the pose has to be retrieved as well as the weights. The results are consistent over various choices of numbers of clusters $M$ (see Supplementary Table S2, and the MCMC convergence diagnostics in Supplementary Table S3). The results are also similar when using a different clustering methodology \cite{klem2022size} (Supplementary Table S4).  In terms of computational costs, one would want to have many cluster centers to have a good representative ensemble, while minimizing the time needed to compute the pose and structure-image likelihood.

\begin{table}[h!]
\centering
\begin{tabular}{|l|c|c|c|}
\hline
 & \%folded          & \%misfolded       & \%unfolded        \\
\hline
Ground   & 0.7707             & 0.0004             & 0.2289           \\
\hline
SNR & \multicolumn{3}{|c|}{ } \\
\hline
% $\infty$ (No noise) & 0.757 $\pm$ 0.002 & 0.0056 $\pm$ 0.0003 & 0.237 $\pm$ 0.002 \\
% 1 & 0.758 $\pm$ 0.002 & 0.0053 $\pm$ 0.0003 & 0.237 $\pm$ 0.002 \\
% 10$^{-1}$ & 0.756 $\pm$ 0.002 & 0.0056 $\pm$ 0.0003 & 0.238 $\pm$ 0.002 \\
% 10$^{-2}$ & 0.742 $\pm$ 0.002 & 0.0123 $\pm$ 0.0004 & 0.246 $\pm$ 0.002 \\
% 10$^{-3}$ & 0.687 $\pm$ 0.002 & 0.036 $\pm$ 0.002 & 0.277 $\pm$ 0.002 \\
% 10$^{-4}$ & 0.549 $\pm$ 0.006 & 0.061 $\pm$ 0.005 & 0.390 $\pm$ 0.005 \\
%%% fixed A and B = 0.5
No noise & 0.758 $\pm$ 0.002 & 0.0049 $\pm$ 0.0003 & 0.237 $\pm$ 0.002 \\
1 & 0.758 $\pm$ 0.002 & 0.0048 $\pm$ 0.0002 & 0.237 $\pm$ 0.002 \\
10$^{-1}$ & 0.757 $\pm$ 0.002 & 0.0049 $\pm$ 0.0003 & 0.238 $\pm$ 0.002 \\
10$^{-2}$ & 0.752 $\pm$ 0.002 & 0.0066 $\pm$ 0.0003 & 0.241 $\pm$ 0.002 \\
10$^{-3}$ & 0.710 $\pm$ 0.002 & 0.022 $\pm$ 0.002 & 0.268 $\pm$ 0.002 \\
10$^{-4}$ & 0.646 $\pm$ 0.007 & 0.047 $\pm$ 0.006 & 0.307 $\pm$ 0.006 \\
\hline
\end{tabular}
\caption{Retrieved populations for the three metastable states of Chignolin. $M=20$ cluster centers ${\ccenter_m}$  are extracted from the structure-generating trajectory, and these are reweighted against the synthetic images with a ground truth population. The expected weight $\alpha_m$ (obtained by sampling Eq.~\ref{eq:final_post_on_clusters}) is assigned to each structure belonging to cluster $m$, and then the retrieved population is the weighted sum of the structures belonging to each state.
The standard deviation of the estimate for the MCMC samples is shown.}
\label{tab:replica_desres_optweights}
\end{table}

\begin{figure}[h!]
\centering
\includegraphics[width=\columnwidth]{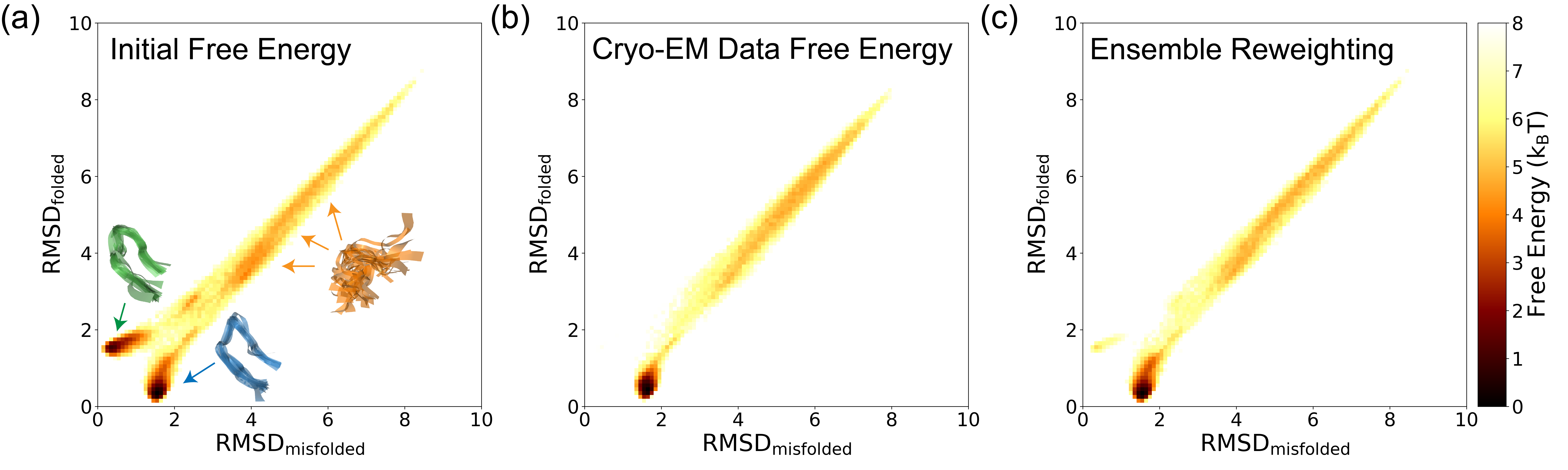}
\caption{Free-energy reweighting. 
2D free-energy surface as a function of the 
 C$_\alpha$ RMSD with respect to the misfolded structure ($x$-axis), and 
 of the C$_\alpha$ RMSD with respect to the folded structure ($y$-axis) for
(A) the initial structure-generating trajectory,
(B) the cryo-EM synthetic images, and
(C) for the structure-generating trajectory reweighted by approximating the ensemble density on clusters (section \ref{ER-cluster}). Reweighting is performed using the image set with SNR$=0.1$.
}
\label{fig:fig4}
\end{figure}

An advantage of working directly in conformational space is that after extracting the cluster center weights, one can then perform post-processing steps to calculate free-energy surfaces as described in the Methods.
In Figure~\ref{fig:fig4}, we show that the approach that approximates the ensemble density on clusters from sampled configurations (section \ref{ER-cluster}) can be used to reweight a 2D free-energy surface. We chose two collective variables: the C$_\alpha$ RMSD with respect to folded and misfolded structures, respectively. The local minima (metastable states) in the surface correspond to darker density regions.
Figure~\ref{fig:fig4}A and B show the initial and ground truth free-energy surface of the structure-generating and image-generating trajectories, respectively. We use the ensemble reweighting on clusters to build a reweighted free-energy surface given the images (Figure~\ref{fig:fig4}C). 
Note that in the initial surface (Figure~\ref{fig:fig4}A) both folded and misfolded states have low free energy. While for the cryo-EM data ensemble (Figure~\ref{fig:fig4}B), the misfolded 
 state has a much higher value (\textit{i.e.}, misfolded conformations are rare). Importantly, this is also found for the reweighted surface (Figure~\ref{fig:fig4}C), where the free-energy minimum of the misfolded state significantly diminishes.
This demonstrates that the less populated misfolded state is down-weighted because of the data and that the reweighted ensemble can be projected onto the low-dimensional free-energy surface. These results show that our algorithm is able to reconstruct free-energy surfaces from cryo-EM images.
The reconstructed free-energy surface is consistent over various choices of $M$ and SNRs (Supplementary Figure S3).  For larger $M$, \textit{e.g.} $M=100$, the free-energy surface better resembles the ground truth.

\section{Conclusions}

We have developed a general ensemble reweighting formalism using single-particle i.i.d.\ observations instead of averaged measurements. Bayesian inference enables retrieving interpretable ensemble weights for a toy model with multiple modes and normally distributed data using a simple reweighting correction (section ref{ER-simple}), even in cases where the ensemble members are not placed at the mode centers. More generally, by leveraging the power of biomolecular modeling and simulations tools, which generate conformational samples, we can reweigh the distribution of clusters from the sampled conformations (section \ref{ER-cluster}). Our results show an accurate metastable population and free-energy recovery using synthetic cryo-EM images with high noise and for complex biomolecules that involve large conformational changes such as unfolded states. The clustering reduces the redundant conformations while maintaining the conformational diversity of the heterogeneous unfolded state. It also simplifies the amount of computation for the image-conformation likelihood, Eq.~\ref{eq:likeli_cryo}, considering that the number of images in a typical cryo-EM particles dataset (and in this study) is in the order of hundreds of thousands. 
The methodology has great potential for the analysis of highly flexible systems having many degrees of freedom (like the unfolded states of biomolecules) using cryo-EM. Importantly, the method provides an approximation of the full Boltzmann ensemble in configuration space without requiring prior collective variables or dimensionality reductions. Moreover, if more conformational samples are generated, one can use the extracted density to evaluate their probability.    

In this work, we validated the method using only synthetic particles of a small protein without estimating the optimal viewing angle for each particle. Analyzing real cryo-EM data will be more challenging because the image-structure likelihood will not be as accurate for high noise levels (where it is difficult to estimate the optimal viewing angle) and the pose optimization will add computational costs. Another limitation is that one requires large computational resources to provide a sufficiently good ensemble to represent all the particle images. For future work, coupling this method with a direct optimization of the conformations (\textit{e.g.}, using MD) would be helpful to efficiently sample the ensemble using the cryo-EM data. The development of more expressive functional forms for $\rwt$ might also be useful.  For instance, one could consider representing $\rwt$ as the output of a neural network or as a draw from a Gaussian process. In conclusion, this work leads the cryo-EM field toward a more quantitative characterization of conformational landscapes. It enables extracting ensemble densities using cryo-EM particles, instead of 3D maps, even for challenging systems where generating reconstructions with cryo-EM is not possible (i.e., highly flexible states). The formalism is sufficiently general that it can be extended to other single-molecule techniques that involve conformational snapshots.

\section*{Author Contributions}

E.H.T, A.B., and P.C. designed the theoretical framework. W.S.T. designed and implemented the computational framework, performed the MD simulations, and analyzed the data. D.S. and J.G.B. contributed to the implementation of the computational framework. 
B.C. wrote the Stan code and revised the manuscript. 
S.H. contributed to the interpretation of the results. 
W.S.T., E.H.T., and P.C. wrote the manuscript. 

\section*{Conflicts of interest}
There are no conflicts to declare.

%\section*{Code availability}

\begin{acknowledgement}
The authors were supported by the Simons Foundation.
\end{acknowledgement}

%\begin{suppinfo}

% \input{SI}

%\end{suppinfo}

%\bibliographystyle
\bibliography{bib}

\providecommand{\latin}[1]{#1}
\makeatletter
\providecommand{\doi}
  {\begingroup\let\do\@makeother\dospecials
  \catcode`\{=1 \catcode`\}=2 \doi@aux}
\providecommand{\doi@aux}[1]{\endgroup\texttt{#1}}
\makeatother
\providecommand*\mcitethebibliography{\thebibliography}
\csname @ifundefined\endcsname{endmcitethebibliography}
  {\let\endmcitethebibliography\endthebibliography}{}
\begin{mcitethebibliography}{56}
\providecommand*\natexlab[1]{#1}
\providecommand*\mciteSetBstSublistMode[1]{}
\providecommand*\mciteSetBstMaxWidthForm[2]{}
\providecommand*\mciteBstWouldAddEndPuncttrue
  {\def\EndOfBibitem{\unskip.}}
\providecommand*\mciteBstWouldAddEndPunctfalse
  {\let\EndOfBibitem\relax}
\providecommand*\mciteSetBstMidEndSepPunct[3]{}
\providecommand*\mciteSetBstSublistLabelBeginEnd[3]{}
\providecommand*\EndOfBibitem{}
\mciteSetBstSublistMode{f}
\mciteSetBstMaxWidthForm{subitem}{(\alph{mcitesubitemcount})}
\mciteSetBstSublistLabelBeginEnd
  {\mcitemaxwidthsubitemform\space}
  {\relax}
  {\relax}

\bibitem[McMullan \latin{et~al.}(2016)McMullan, Faruqi, and
  Henderson]{mcmullan2016direct}
McMullan,~G.; Faruqi,~A.; Henderson,~R. Direct electron detectors.
  \emph{Methods in enzymology} \textbf{2016}, \emph{579}, 1--17\relax
\mciteBstWouldAddEndPuncttrue
\mciteSetBstMidEndSepPunct{\mcitedefaultmidpunct}
{\mcitedefaultendpunct}{\mcitedefaultseppunct}\relax
\EndOfBibitem
\bibitem[Li \latin{et~al.}(2013)Li, Mooney, Zheng, Booth, Braunfeld, Gubbens,
  Agard, and Cheng]{li2013electron}
Li,~X.; Mooney,~P.; Zheng,~S.; Booth,~C.~R.; Braunfeld,~M.~B.; Gubbens,~S.;
  Agard,~D.~A.; Cheng,~Y. Electron counting and beam-induced motion correction
  enable near-atomic-resolution single-particle cryo-{EM}. \emph{Nature
  methods} \textbf{2013}, \emph{10}, 584--590\relax
\mciteBstWouldAddEndPuncttrue
\mciteSetBstMidEndSepPunct{\mcitedefaultmidpunct}
{\mcitedefaultendpunct}{\mcitedefaultseppunct}\relax
\EndOfBibitem
\bibitem[Cossio and Hummer(2018)Cossio, and Hummer]{cossio2018likelihood}
Cossio,~P.; Hummer,~G. Likelihood-based structural analysis of electron
  microscopy images. \emph{Current opinion in structural biology}
  \textbf{2018}, \emph{49}, 162--168\relax
\mciteBstWouldAddEndPuncttrue
\mciteSetBstMidEndSepPunct{\mcitedefaultmidpunct}
{\mcitedefaultendpunct}{\mcitedefaultseppunct}\relax
\EndOfBibitem
\bibitem[K{\"u}hlbrandt(2014)]{kuhlbrandt2014resolution}
K{\"u}hlbrandt,~W. The resolution revolution. \emph{Science} \textbf{2014},
  \emph{343}, 1443--1444\relax
\mciteBstWouldAddEndPuncttrue
\mciteSetBstMidEndSepPunct{\mcitedefaultmidpunct}
{\mcitedefaultendpunct}{\mcitedefaultseppunct}\relax
\EndOfBibitem
\bibitem[Nakane \latin{et~al.}(2020)Nakane, Kotecha, Sente, McMullan, Masiulis,
  Brown, Grigoras, Malinauskaite, Malinauskas, Miehling, Uchański, Yu, Karia,
  Pechnikova, de~Jong, Keizer, Bischoff, McCormack, Tiemeijer, Hardwick,
  Chirgadze, Murshudov, Aricescu, and Scheres]{Nakane2020}
Nakane,~T. \latin{et~al.}  Single-particle cryo-{EM} at atomic resolution.
  \emph{Nature} \textbf{2020}, \emph{587}, 152--156\relax
\mciteBstWouldAddEndPuncttrue
\mciteSetBstMidEndSepPunct{\mcitedefaultmidpunct}
{\mcitedefaultendpunct}{\mcitedefaultseppunct}\relax
\EndOfBibitem
\bibitem[Ward \latin{et~al.}(2013)Ward, Sali, and Wilson]{ward2013integrative}
Ward,~A.~B.; Sali,~A.; Wilson,~I.~A. Integrative structural biology.
  \emph{Science} \textbf{2013}, \emph{339}, 913--915\relax
\mciteBstWouldAddEndPuncttrue
\mciteSetBstMidEndSepPunct{\mcitedefaultmidpunct}
{\mcitedefaultendpunct}{\mcitedefaultseppunct}\relax
\EndOfBibitem
\bibitem[Bottaro and Lindorff-Larsen(2018)Bottaro, and
  Lindorff-Larsen]{bottaro2018biophysical}
Bottaro,~S.; Lindorff-Larsen,~K. Biophysical experiments and biomolecular
  simulations: A perfect match? \emph{Science} \textbf{2018}, \emph{361},
  355--360\relax
\mciteBstWouldAddEndPuncttrue
\mciteSetBstMidEndSepPunct{\mcitedefaultmidpunct}
{\mcitedefaultendpunct}{\mcitedefaultseppunct}\relax
\EndOfBibitem
\bibitem[Bonomi \latin{et~al.}(2017)Bonomi, Heller, Camilloni, and
  Vendruscolo]{bonomi2017principles}
Bonomi,~M.; Heller,~G.~T.; Camilloni,~C.; Vendruscolo,~M. Principles of protein
  structural ensemble determination. \emph{Current opinion in structural
  biology} \textbf{2017}, \emph{42}, 106--116\relax
\mciteBstWouldAddEndPuncttrue
\mciteSetBstMidEndSepPunct{\mcitedefaultmidpunct}
{\mcitedefaultendpunct}{\mcitedefaultseppunct}\relax
\EndOfBibitem
\bibitem[Costa and Fushman(2022)Costa, and Fushman]{costa2022reweighting}
Costa,~R. G.~L.; Fushman,~D. Reweighting methods for elucidation of
  conformation ensembles of proteins. \emph{Current Opinion in Structural
  Biology} \textbf{2022}, \emph{77}, 102470\relax
\mciteBstWouldAddEndPuncttrue
\mciteSetBstMidEndSepPunct{\mcitedefaultmidpunct}
{\mcitedefaultendpunct}{\mcitedefaultseppunct}\relax
\EndOfBibitem
\bibitem[Rieping \latin{et~al.}(2005)Rieping, Habeck, and
  Nilges]{rieping2005inferential}
Rieping,~W.; Habeck,~M.; Nilges,~M. Inferential structure determination.
  \emph{Science} \textbf{2005}, \emph{309}, 303--306\relax
\mciteBstWouldAddEndPuncttrue
\mciteSetBstMidEndSepPunct{\mcitedefaultmidpunct}
{\mcitedefaultendpunct}{\mcitedefaultseppunct}\relax
\EndOfBibitem
\bibitem[R{\'o}{\.z}ycki \latin{et~al.}(2011)R{\'o}{\.z}ycki, Kim, and
  Hummer]{rozycki2011saxs}
R{\'o}{\.z}ycki,~B.; Kim,~Y.~C.; Hummer,~G. SAXS ensemble refinement of
  ESCRT-III CHMP3 conformational transitions. \emph{Structure} \textbf{2011},
  \emph{19}, 109--116\relax
\mciteBstWouldAddEndPuncttrue
\mciteSetBstMidEndSepPunct{\mcitedefaultmidpunct}
{\mcitedefaultendpunct}{\mcitedefaultseppunct}\relax
\EndOfBibitem
\bibitem[Boomsma \latin{et~al.}(2014)Boomsma, Ferkinghoff-Borg, and
  Lindorff-Larsen]{boomsma2014combining}
Boomsma,~W.; Ferkinghoff-Borg,~J.; Lindorff-Larsen,~K. Combining experiments
  and simulations using the maximum entropy principle. \emph{PLoS computational
  biology} \textbf{2014}, \emph{10}, e1003406\relax
\mciteBstWouldAddEndPuncttrue
\mciteSetBstMidEndSepPunct{\mcitedefaultmidpunct}
{\mcitedefaultendpunct}{\mcitedefaultseppunct}\relax
\EndOfBibitem
\bibitem[Hummer and K{\"o}finger(2015)Hummer, and
  K{\"o}finger]{hummer2015bayesian}
Hummer,~G.; K{\"o}finger,~J. Bayesian ensemble refinement by replica
  simulations and reweighting. \emph{The Journal of chemical physics}
  \textbf{2015}, \emph{143}, 12B634\_1\relax
\mciteBstWouldAddEndPuncttrue
\mciteSetBstMidEndSepPunct{\mcitedefaultmidpunct}
{\mcitedefaultendpunct}{\mcitedefaultseppunct}\relax
\EndOfBibitem
\bibitem[K{\"o}finger \latin{et~al.}(2019)K{\"o}finger, Stelzl, Reuter,
  Allande, Reichel, and Hummer]{kofinger2019efficient}
K{\"o}finger,~J.; Stelzl,~L.~S.; Reuter,~K.; Allande,~C.; Reichel,~K.;
  Hummer,~G. Efficient ensemble refinement by reweighting. \emph{Journal of
  chemical theory and computation} \textbf{2019}, \emph{15}, 3390--3401\relax
\mciteBstWouldAddEndPuncttrue
\mciteSetBstMidEndSepPunct{\mcitedefaultmidpunct}
{\mcitedefaultendpunct}{\mcitedefaultseppunct}\relax
\EndOfBibitem
\bibitem[Bonomi \latin{et~al.}(2016)Bonomi, Camilloni, Cavalli, and
  Vendruscolo]{bonomi2016metainference}
Bonomi,~M.; Camilloni,~C.; Cavalli,~A.; Vendruscolo,~M. Metainference: A
  Bayesian inference method for heterogeneous systems. \emph{Science advances}
  \textbf{2016}, \emph{2}, e1501177\relax
\mciteBstWouldAddEndPuncttrue
\mciteSetBstMidEndSepPunct{\mcitedefaultmidpunct}
{\mcitedefaultendpunct}{\mcitedefaultseppunct}\relax
\EndOfBibitem
\bibitem[Barrett \latin{et~al.}(2022)Barrett, Ansari, Ghoshal, and
  White]{barrett2022simulation}
Barrett,~R.; Ansari,~M.; Ghoshal,~G.; White,~A.~D. Simulation-based inference
  with approximately correct parameters via maximum entropy. \emph{Machine
  Learning: Science and Technology} \textbf{2022}, \emph{3}, 025006\relax
\mciteBstWouldAddEndPuncttrue
\mciteSetBstMidEndSepPunct{\mcitedefaultmidpunct}
{\mcitedefaultendpunct}{\mcitedefaultseppunct}\relax
\EndOfBibitem
\bibitem[Roux and Weare(2013)Roux, and Weare]{roux2013statistical}
Roux,~B.; Weare,~J. On the statistical equivalence of restrained-ensemble
  simulations with the maximum entropy method. \emph{The Journal of chemical
  physics} \textbf{2013}, \emph{138}, 02B616\relax
\mciteBstWouldAddEndPuncttrue
\mciteSetBstMidEndSepPunct{\mcitedefaultmidpunct}
{\mcitedefaultendpunct}{\mcitedefaultseppunct}\relax
\EndOfBibitem
\bibitem[Cesari \latin{et~al.}(2018)Cesari, Rei{\ss}er, and
  Bussi]{cesari2018using}
Cesari,~A.; Rei{\ss}er,~S.; Bussi,~G. Using the maximum entropy principle to
  combine simulations and solution experiments. \emph{Computation}
  \textbf{2018}, \emph{6}, 15\relax
\mciteBstWouldAddEndPuncttrue
\mciteSetBstMidEndSepPunct{\mcitedefaultmidpunct}
{\mcitedefaultendpunct}{\mcitedefaultseppunct}\relax
\EndOfBibitem
\bibitem[Trabuco \latin{et~al.}(2009)Trabuco, Villa, Schreiner, Harrison, and
  Schulten]{trabuco2009molecular}
Trabuco,~L.~G.; Villa,~E.; Schreiner,~E.; Harrison,~C.~B.; Schulten,~K.
  Molecular dynamics flexible fitting: a practical guide to combine
  cryo-electron microscopy and X-ray crystallography. \emph{Methods}
  \textbf{2009}, \emph{49}, 174--180\relax
\mciteBstWouldAddEndPuncttrue
\mciteSetBstMidEndSepPunct{\mcitedefaultmidpunct}
{\mcitedefaultendpunct}{\mcitedefaultseppunct}\relax
\EndOfBibitem
\bibitem[Igaev \latin{et~al.}(2019)Igaev, Kutzner, Bock, Vaiana, and
  Grubm{\"u}ller]{igaev2019automated}
Igaev,~M.; Kutzner,~C.; Bock,~L.~V.; Vaiana,~A.~C.; Grubm{\"u}ller,~H.
  Automated cryo-{EM} structure refinement using correlation-driven molecular
  dynamics. \emph{Elife} \textbf{2019}, \emph{8}, e43542\relax
\mciteBstWouldAddEndPuncttrue
\mciteSetBstMidEndSepPunct{\mcitedefaultmidpunct}
{\mcitedefaultendpunct}{\mcitedefaultseppunct}\relax
\EndOfBibitem
\bibitem[Mori \latin{et~al.}(2021)Mori, Terashi, Matsuoka, Kihara, and
  Sugita]{mori2021efficient}
Mori,~T.; Terashi,~G.; Matsuoka,~D.; Kihara,~D.; Sugita,~Y. Efficient Flexible
  Fitting Refinement with Automatic Error Fixing for De Novo Structure Modeling
  from Cryo-{EM} Density Maps. \emph{Journal of Chemical Information and
  Modeling} \textbf{2021}, \emph{61}, 3516--3528\relax
\mciteBstWouldAddEndPuncttrue
\mciteSetBstMidEndSepPunct{\mcitedefaultmidpunct}
{\mcitedefaultendpunct}{\mcitedefaultseppunct}\relax
\EndOfBibitem
\bibitem[Blau \latin{et~al.}(2022)Blau, Yvonnesdotter, and
  Lindahl]{blau2022gentle}
Blau,~C.; Yvonnesdotter,~L.; Lindahl,~E. Gentle and fast all-atom model
  refinement to cryo-{EM} densities via {B}ayes' approach. \emph{bioRxiv}
  \textbf{2022}, \relax
\mciteBstWouldAddEndPunctfalse
\mciteSetBstMidEndSepPunct{\mcitedefaultmidpunct}
{}{\mcitedefaultseppunct}\relax
\EndOfBibitem
\bibitem[Vuillemot \latin{et~al.}(2022)Vuillemot, Miyashita, Tama, Rouiller,
  and Jonic]{vuillemot2022nmmd}
Vuillemot,~R.; Miyashita,~O.; Tama,~F.; Rouiller,~I.; Jonic,~S. NMMD: Efficient
  cryo-{EM} flexible fitting based on simultaneous Normal Mode and Molecular
  Dynamics atomic displacements. \emph{Journal of Molecular Biology}
  \textbf{2022}, \emph{434}, 167483\relax
\mciteBstWouldAddEndPuncttrue
\mciteSetBstMidEndSepPunct{\mcitedefaultmidpunct}
{\mcitedefaultendpunct}{\mcitedefaultseppunct}\relax
\EndOfBibitem
\bibitem[Bonomi \latin{et~al.}(2018)Bonomi, Pellarin, and
  Vendruscolo]{bonomi2018simultaneous}
Bonomi,~M.; Pellarin,~R.; Vendruscolo,~M. Simultaneous determination of protein
  structure and dynamics using cryo-electron microscopy. \emph{Biophysical
  journal} \textbf{2018}, \emph{114}, 1604--1613\relax
\mciteBstWouldAddEndPuncttrue
\mciteSetBstMidEndSepPunct{\mcitedefaultmidpunct}
{\mcitedefaultendpunct}{\mcitedefaultseppunct}\relax
\EndOfBibitem
\bibitem[Dubochet \latin{et~al.}(1988)Dubochet, Adrian, Chang, Homo, Lepault,
  McDowall, and Schultz]{Dubochet1988}
Dubochet,~J.; Adrian,~M.; Chang,~J.-J.; Homo,~J.-C.; Lepault,~J.;
  McDowall,~A.~W.; Schultz,~P. {Cryo-electron microscopy of vitrified
  specimens}. \emph{Q. Rev. Biophys.} \textbf{1988}, \emph{21}, 129--228\relax
\mciteBstWouldAddEndPuncttrue
\mciteSetBstMidEndSepPunct{\mcitedefaultmidpunct}
{\mcitedefaultendpunct}{\mcitedefaultseppunct}\relax
\EndOfBibitem
\bibitem[Fischer \latin{et~al.}(2010)Fischer, Konevega, Wintermeyer, Rodnina,
  and Stark]{Fischer2010}
Fischer,~N.; Konevega,~A.~L.; Wintermeyer,~W.; Rodnina,~M.~V.; Stark,~H.
  {Ribosome dynamics and tRNA movement by time-resolved electron
  cryomicroscopy}. \emph{Nature} \textbf{2010}, \emph{466}, 329--333\relax
\mciteBstWouldAddEndPuncttrue
\mciteSetBstMidEndSepPunct{\mcitedefaultmidpunct}
{\mcitedefaultendpunct}{\mcitedefaultseppunct}\relax
\EndOfBibitem
\bibitem[Dashti \latin{et~al.}(2014)Dashti, Schwander, Langlois, Fung, Li,
  Hosseinizadeh, Liao, Pallesen, Sharma, Stupina, Simon, Dinman, Frank, and
  Ourmazd]{Dashti2014}
Dashti,~A.; Schwander,~P.; Langlois,~R.; Fung,~R.; Li,~W.; Hosseinizadeh,~A.;
  Liao,~H.~Y.; Pallesen,~J.; Sharma,~G.; Stupina,~V.~A.; Simon,~A.~E.;
  Dinman,~J.~D.; Frank,~J.; Ourmazd,~A. {Trajectories of the ribosome as a
  Brownian nanomachine}. \emph{Proc. Natl. Acad. Sci. U. S. A.} \textbf{2014},
  \emph{111}, 17492--17497\relax
\mciteBstWouldAddEndPuncttrue
\mciteSetBstMidEndSepPunct{\mcitedefaultmidpunct}
{\mcitedefaultendpunct}{\mcitedefaultseppunct}\relax
\EndOfBibitem
\bibitem[Dashti \latin{et~al.}(2020)Dashti, Mashayekhi, Shekhar, {Ben Hail},
  Salah, Schwander, des Georges, Singharoy, Frank, and Ourmazd]{Dashti2020}
Dashti,~A.; Mashayekhi,~G.; Shekhar,~M.; {Ben Hail},~D.; Salah,~S.;
  Schwander,~P.; des Georges,~A.; Singharoy,~A.; Frank,~J.; Ourmazd,~A.
  {Retrieving functional pathways of biomolecules from single-particle
  snapshots}. \emph{Nat. Commun.} \textbf{2020}, \emph{11}, 4734\relax
\mciteBstWouldAddEndPuncttrue
\mciteSetBstMidEndSepPunct{\mcitedefaultmidpunct}
{\mcitedefaultendpunct}{\mcitedefaultseppunct}\relax
\EndOfBibitem
\bibitem[Cossio and Hummer(2013)Cossio, and Hummer]{cossio2013bayesian}
Cossio,~P.; Hummer,~G. {Bayesian analysis of individual electron microscopy
  images: Towards structures of dynamic and heterogeneous biomolecular
  assemblies}. \emph{Journal of Structural Biology} \textbf{2013}, \emph{184},
  427--437\relax
\mciteBstWouldAddEndPuncttrue
\mciteSetBstMidEndSepPunct{\mcitedefaultmidpunct}
{\mcitedefaultendpunct}{\mcitedefaultseppunct}\relax
\EndOfBibitem
\bibitem[Giraldo-Barreto \latin{et~al.}(2021)Giraldo-Barreto, Ortiz, Thiede,
  Palacio-Rodriguez, Carpenter, Barnett, and Cossio]{giraldo2021bayesian}
Giraldo-Barreto,~J.; Ortiz,~S.; Thiede,~E.~H.; Palacio-Rodriguez,~K.;
  Carpenter,~B.; Barnett,~A.~H.; Cossio,~P. A Bayesian approach to extracting
  free-energy profiles from cryo-electron microscopy experiments.
  \emph{Scientific Reports} \textbf{2021}, \emph{11}, 1--15\relax
\mciteBstWouldAddEndPuncttrue
\mciteSetBstMidEndSepPunct{\mcitedefaultmidpunct}
{\mcitedefaultendpunct}{\mcitedefaultseppunct}\relax
\EndOfBibitem
\bibitem[Latorraca \latin{et~al.}(2017)Latorraca, Fastman, Venkatakrishnan,
  Frommer, Dror, and Feng]{latorraca2017mechanism}
Latorraca,~N.~R.; Fastman,~N.~M.; Venkatakrishnan,~A.; Frommer,~W.~B.;
  Dror,~R.~O.; Feng,~L. Mechanism of substrate translocation in an alternating
  access transporter. \emph{Cell} \textbf{2017}, \emph{169}, 96--107\relax
\mciteBstWouldAddEndPuncttrue
\mciteSetBstMidEndSepPunct{\mcitedefaultmidpunct}
{\mcitedefaultendpunct}{\mcitedefaultseppunct}\relax
\EndOfBibitem
\bibitem[Frenkel and Smit(2002)Frenkel, and Smit]{FrenkelSmit2002}
Frenkel,~D.; Smit,~B. \emph{Understanding molecular simulation from algorithms
  to applications}; Computational Science; Academic Press: San Diego, 2002; p
  638\relax
\mciteBstWouldAddEndPuncttrue
\mciteSetBstMidEndSepPunct{\mcitedefaultmidpunct}
{\mcitedefaultendpunct}{\mcitedefaultseppunct}\relax
\EndOfBibitem
\bibitem[Das and Baker(2008)Das, and Baker]{das2008macromolecular}
Das,~R.; Baker,~D. Macromolecular modeling with {R}osetta. \emph{Annual Revew
  of Biochemistry} \textbf{2008}, \emph{77}, 363--382\relax
\mciteBstWouldAddEndPuncttrue
\mciteSetBstMidEndSepPunct{\mcitedefaultmidpunct}
{\mcitedefaultendpunct}{\mcitedefaultseppunct}\relax
\EndOfBibitem
\bibitem[Leman \latin{et~al.}(2020)Leman, Weitzner, Lewis, Adolf-Bryfogle,
  Alam, Alford, Aprahamian, Baker, Barlow, Barth, \latin{et~al.}
  others]{leman2020macromolecular}
Leman,~J.~K.; Weitzner,~B.~D.; Lewis,~S.~M.; Adolf-Bryfogle,~J.; Alam,~N.;
  Alford,~R.~F.; Aprahamian,~M.; Baker,~D.; Barlow,~K.~A.; Barth,~P.,
  \latin{et~al.}  Macromolecular modeling and design in {R}osetta: recent
  methods and frameworks. \emph{Nature Methods} \textbf{2020}, \emph{17},
  665--680\relax
\mciteBstWouldAddEndPuncttrue
\mciteSetBstMidEndSepPunct{\mcitedefaultmidpunct}
{\mcitedefaultendpunct}{\mcitedefaultseppunct}\relax
\EndOfBibitem
\bibitem[No{\'e} \latin{et~al.}(2019)No{\'e}, Olsson, K{\"o}hler, and
  Wu]{noe2019boltzmann}
No{\'e},~F.; Olsson,~S.; K{\"o}hler,~J.; Wu,~H. Boltzmann generators:
  {S}ampling equilibrium states of many-body systems with deep learning.
  \emph{Science} \textbf{2019}, \emph{365}, eaaw1147\relax
\mciteBstWouldAddEndPuncttrue
\mciteSetBstMidEndSepPunct{\mcitedefaultmidpunct}
{\mcitedefaultendpunct}{\mcitedefaultseppunct}\relax
\EndOfBibitem
\bibitem[Degiacomi(2019)]{degiacomi2019coupling}
Degiacomi,~M.~T. Coupling molecular dynamics and deep learning to mine protein
  conformational space. \emph{Structure} \textbf{2019}, \emph{27},
  1034--1040\relax
\mciteBstWouldAddEndPuncttrue
\mciteSetBstMidEndSepPunct{\mcitedefaultmidpunct}
{\mcitedefaultendpunct}{\mcitedefaultseppunct}\relax
\EndOfBibitem
\bibitem[Vani \latin{et~al.}(2022)Vani, Aranganathan, Wang, and
  Tiwary]{vani2022sequence}
Vani,~B.~P.; Aranganathan,~A.; Wang,~D.; Tiwary,~P. From sequence to
  {B}oltzmann weighted ensemble of structures with {A}lpha{F}old2-{RAVE}.
  \emph{bioRxiv} \textbf{2022}, \relax
\mciteBstWouldAddEndPunctfalse
\mciteSetBstMidEndSepPunct{\mcitedefaultmidpunct}
{}{\mcitedefaultseppunct}\relax
\EndOfBibitem
\bibitem[Hoffman \latin{et~al.}(2014)Hoffman, Gelman, \latin{et~al.}
  others]{hoffman2014nuts}
Hoffman,~M.~D.; Gelman,~A., \latin{et~al.}  The {No-U-Turn} sampler: adaptively
  setting path lengths in {Hamiltonian Monte Carlo}. \emph{J. Mach. Learn.
  Res.} \textbf{2014}, \emph{15}, 1593--1623\relax
\mciteBstWouldAddEndPuncttrue
\mciteSetBstMidEndSepPunct{\mcitedefaultmidpunct}
{\mcitedefaultendpunct}{\mcitedefaultseppunct}\relax
\EndOfBibitem
\bibitem[Carpenter \latin{et~al.}(2017)Carpenter, Gelman, Hoffman, Lee,
  Goodrich, Betancourt, Brubaker, Guo, Li, and Riddell]{carpenter2017stan}
Carpenter,~B.; Gelman,~A.; Hoffman,~M.~D.; Lee,~D.; Goodrich,~B.;
  Betancourt,~M.; Brubaker,~M.; Guo,~J.; Li,~P.; Riddell,~A. Stan: A
  probabilistic programming language. \emph{Journal of statistical software}
  \textbf{2017}, \emph{76}\relax
\mciteBstWouldAddEndPuncttrue
\mciteSetBstMidEndSepPunct{\mcitedefaultmidpunct}
{\mcitedefaultendpunct}{\mcitedefaultseppunct}\relax
\EndOfBibitem
\bibitem[Seitz \latin{et~al.}(2019)Seitz, Acosta-Reyes, Schwander, and
  Frank]{seitz2019simulation}
Seitz,~E.; Acosta-Reyes,~F.; Schwander,~P.; Frank,~J. Simulation of cryo-{EM}
  ensembles from atomic models of molecules exhibiting continuous
  conformations. \emph{BioRxiv} \textbf{2019}, 864116\relax
\mciteBstWouldAddEndPuncttrue
\mciteSetBstMidEndSepPunct{\mcitedefaultmidpunct}
{\mcitedefaultendpunct}{\mcitedefaultseppunct}\relax
\EndOfBibitem
\bibitem[Bendory \latin{et~al.}(2020)Bendory, Bartesaghi, and
  Singer]{bendory2020single}
Bendory,~T.; Bartesaghi,~A.; Singer,~A. Single-particle cryo-electron
  microscopy: Mathematical theory, computational challenges, and opportunities.
  \emph{IEEE signal processing magazine} \textbf{2020}, \emph{37}, 58--76\relax
\mciteBstWouldAddEndPuncttrue
\mciteSetBstMidEndSepPunct{\mcitedefaultmidpunct}
{\mcitedefaultendpunct}{\mcitedefaultseppunct}\relax
\EndOfBibitem
\bibitem[Penczek(2010)]{grant2010image}
Penczek,~P.~A. In \emph{Cryo-{EM}, Part B: 3-D Reconstruction}; Jensen,~G.~J.,
  Ed.; Methods in Enzymology; Academic Press, 2010; Vol. 482; pp 35--72\relax
\mciteBstWouldAddEndPuncttrue
\mciteSetBstMidEndSepPunct{\mcitedefaultmidpunct}
{\mcitedefaultendpunct}{\mcitedefaultseppunct}\relax
\EndOfBibitem
\bibitem[Cossio \latin{et~al.}(2017)Cossio, Rohr, Baruffa, Rampp, Lindenstruth,
  and Hummer]{cossio2017bioem}
Cossio,~P.; Rohr,~D.; Baruffa,~F.; Rampp,~M.; Lindenstruth,~V.; Hummer,~G.
  BioEM: GPU-accelerated computing of Bayesian inference of electron microscopy
  images. \emph{Computer Physics Communications} \textbf{2017}, \emph{210},
  163--171\relax
\mciteBstWouldAddEndPuncttrue
\mciteSetBstMidEndSepPunct{\mcitedefaultmidpunct}
{\mcitedefaultendpunct}{\mcitedefaultseppunct}\relax
\EndOfBibitem
\bibitem[Cossio \latin{et~al.}({2018})Cossio, Allegretti, Mayer, Mueller,
  Vonck, and Hummer]{Cossio:Micro:2018}
Cossio,~P.; Allegretti,~M.; Mayer,~F.; Mueller,~V.; Vonck,~J.; Hummer,~G.
  {Bayesian inference of rotor ring stoichiometry from electron microscopy
  images of archaeal ATP synthase}. \emph{{Microscopy}} \textbf{{2018}},
  \emph{{67}}, {266--273}\relax
\mciteBstWouldAddEndPuncttrue
\mciteSetBstMidEndSepPunct{\mcitedefaultmidpunct}
{\mcitedefaultendpunct}{\mcitedefaultseppunct}\relax
\EndOfBibitem
\bibitem[Lindorff-Larsen \latin{et~al.}(2010)Lindorff-Larsen, Piana, Palmo,
  Maragakis, Klepeis, Dror, and Shaw]{lindorff2010improved}
Lindorff-Larsen,~K.; Piana,~S.; Palmo,~K.; Maragakis,~P.; Klepeis,~J.~L.;
  Dror,~R.~O.; Shaw,~D.~E. Improved side-chain torsion potentials for the
  {A}mber ff99{SB} protein force field. \emph{Proteins: Structure, Function,
  and Bioinformatics} \textbf{2010}, \emph{78}, 1950--1958\relax
\mciteBstWouldAddEndPuncttrue
\mciteSetBstMidEndSepPunct{\mcitedefaultmidpunct}
{\mcitedefaultendpunct}{\mcitedefaultseppunct}\relax
\EndOfBibitem
\bibitem[Abraham \latin{et~al.}(2015)Abraham, Murtola, Schulz, P{\'a}ll, Smith,
  Hess, and Lindahl]{abraham2015gromacs}
Abraham,~M.~J.; Murtola,~T.; Schulz,~R.; P{\'a}ll,~S.; Smith,~J.~C.; Hess,~B.;
  Lindahl,~E. {GROMACS}: High performance molecular simulations through
  multi-level parallelism from laptops to supercomputers. \emph{SoftwareX}
  \textbf{2015}, \emph{1}, 19--25\relax
\mciteBstWouldAddEndPuncttrue
\mciteSetBstMidEndSepPunct{\mcitedefaultmidpunct}
{\mcitedefaultendpunct}{\mcitedefaultseppunct}\relax
\EndOfBibitem
\bibitem[Honda \latin{et~al.}(2004)Honda, Yamasaki, Sawada, and
  Morii]{honda2004chignolin}
Honda,~S.; Yamasaki,~K.; Sawada,~Y.; Morii,~H. 10 residue folded peptide
  designed by segment statistics. \emph{Structure} \textbf{2004}, \emph{12},
  1507--1518\relax
\mciteBstWouldAddEndPuncttrue
\mciteSetBstMidEndSepPunct{\mcitedefaultmidpunct}
{\mcitedefaultendpunct}{\mcitedefaultseppunct}\relax
\EndOfBibitem
\bibitem[Lawrence and Skinner(2003)Lawrence, and Skinner]{lawrence2003tip4p}
Lawrence,~C.; Skinner,~J. Flexible TIP4P model for molecular dynamics
  simulation of liquid water. \emph{Chemical physics letters} \textbf{2003},
  \emph{372}, 842--847\relax
\mciteBstWouldAddEndPuncttrue
\mciteSetBstMidEndSepPunct{\mcitedefaultmidpunct}
{\mcitedefaultendpunct}{\mcitedefaultseppunct}\relax
\EndOfBibitem
\bibitem[Hess \latin{et~al.}(1997)Hess, Bekker, Berendsen, and
  Fraaije]{hess1997lincs}
Hess,~B.; Bekker,~H.; Berendsen,~H.~J.; Fraaije,~J.~G. {LINCS}: a linear
  constraint solver for molecular simulations. \emph{Journal of computational
  chemistry} \textbf{1997}, \emph{18}, 1463--1472\relax
\mciteBstWouldAddEndPuncttrue
\mciteSetBstMidEndSepPunct{\mcitedefaultmidpunct}
{\mcitedefaultendpunct}{\mcitedefaultseppunct}\relax
\EndOfBibitem
\bibitem[Schubert and Lenssen(2022)Schubert, and Lenssen]{schubert2022kmedoids}
Schubert,~E.; Lenssen,~L. Fast k-medoids Clustering in Rust and Python.
  \emph{Journal of Open Source Software} \textbf{2022}, \emph{7}, 4183\relax
\mciteBstWouldAddEndPuncttrue
\mciteSetBstMidEndSepPunct{\mcitedefaultmidpunct}
{\mcitedefaultendpunct}{\mcitedefaultseppunct}\relax
\EndOfBibitem
\bibitem[Schubert and Rousseeuw(2021)Schubert, and Rousseeuw]{schubert2021fast}
Schubert,~E.; Rousseeuw,~P.~J. Fast and eager k-medoids clustering: O (k)
  runtime improvement of the PAM, CLARA, and CLARANS algorithms.
  \emph{Information Systems} \textbf{2021}, \emph{101}, 101804\relax
\mciteBstWouldAddEndPuncttrue
\mciteSetBstMidEndSepPunct{\mcitedefaultmidpunct}
{\mcitedefaultendpunct}{\mcitedefaultseppunct}\relax
\EndOfBibitem
\bibitem[Klem \latin{et~al.}(2022)Klem, Hocky, and McCullagh]{klem2022size}
Klem,~H.; Hocky,~G.~M.; McCullagh,~M. Size-and-Shape Space Gaussian Mixture
  Models for Structural Clustering of Molecular Dynamics Trajectories.
  \emph{Journal of chemical theory and computation} \textbf{2022}, \relax
\mciteBstWouldAddEndPunctfalse
\mciteSetBstMidEndSepPunct{\mcitedefaultmidpunct}
{}{\mcitedefaultseppunct}\relax
\EndOfBibitem
\bibitem[Lindorff-Larsen \latin{et~al.}(2011)Lindorff-Larsen, Piana, Dror, and
  Shaw]{kresten2011how}
Lindorff-Larsen,~K.; Piana,~S.; Dror,~R.~O.; Shaw,~D.~E. How Fast-Folding
  Proteins Fold. \emph{Science} \textbf{2011}, \emph{334}, 517--520\relax
\mciteBstWouldAddEndPuncttrue
\mciteSetBstMidEndSepPunct{\mcitedefaultmidpunct}
{\mcitedefaultendpunct}{\mcitedefaultseppunct}\relax
\EndOfBibitem
\bibitem[Piana \latin{et~al.}(2011)Piana, Lindorff-Larsen, and
  Shaw]{piana2011robust}
Piana,~S.; Lindorff-Larsen,~K.; Shaw,~D.~E. How robust are protein folding
  simulations with respect to force field parameterization? \emph{Biophysical
  journal} \textbf{2011}, \emph{100}, L47--L49\relax
\mciteBstWouldAddEndPuncttrue
\mciteSetBstMidEndSepPunct{\mcitedefaultmidpunct}
{\mcitedefaultendpunct}{\mcitedefaultseppunct}\relax
\EndOfBibitem
\bibitem[Satoh \latin{et~al.}(2006)Satoh, Shimizu, Nakamura, and
  Terada]{satoh2006folding}
Satoh,~D.; Shimizu,~K.; Nakamura,~S.; Terada,~T. Folding free-energy landscape
  of a 10-residue mini-protein, chignolin. \emph{FEBS letters} \textbf{2006},
  \emph{580}, 3422--3426\relax
\mciteBstWouldAddEndPuncttrue
\mciteSetBstMidEndSepPunct{\mcitedefaultmidpunct}
{\mcitedefaultendpunct}{\mcitedefaultseppunct}\relax
\EndOfBibitem
\end{mcitethebibliography}

\end{document}